\documentclass[12pt,a4paper]{article}

\usepackage{amsmath}
\usepackage{caption}
\usepackage{subcaption}
\usepackage{indent first}
\usepackage{jheppub}
\usepackage{pdfcomment}
\usepackage{xspace}
\usepackage{tikz}
\usetikzlibrary{shapes,matrix}
\tikzset{>=latex}
\tikzset{baseline=(A.base)}
\tikzset{flavor/.style={draw}}
\tikzset{gauge/.style={draw,circle,inner sep=2pt}}
\tikzset{T/.style={draw,shape=isosceles triangle,isosceles triangle apex angle=60,inner sep=2pt}}

\usepackage{framed}
\definecolor{shadecolor}{rgb}{0.90,0.90,0.90}
\newenvironment{claim}{\begin{shaded}\noindent\itshape\ignorespaces}{\end{shaded}}

\newcommand{\bC}{\mathbb{C}}
\newcommand{\bR}{\mathbb{R}}
\newcommand{\bZ}{\mathbb{Z}}
\def\Nequals#1{$\mathcal{N}{=}#1$}
\def\SU{\mathrm{SU}}

\def\U{\mathrm{U}}
\def\SO{\mathrm{SO}}
\def\SL{\mathrm{SL}}
\def\tr{\mathop{\mathrm{tr}}\nolimits}
\def\diag{\mathop{\mathrm{diag}}\nolimits}
\def\rank{\mathop{\mathrm{rank}}\nolimits}

\def\Tr{\mathop{\mathrm{Tr}}\nolimits}

\def\vev#1{\langle#1\rangle}
\def\cH{{\cal H}}
\def\ff{\mathfrak{f}}
\def\fg{\mathfrak{g}}
\def\ii{\mathrm{i}}
\def\CP{\mathbb{CP}}
\def\sB{\mathsf{B}}
\def\sS{\mathsf{S}}
\def\sT{\mathsf{T}}
\def\CCt{\mathcal{C}_\mathrm{T}}

\def\beq#1\eeq{\begin{align}#1\end{align}}

\def\CC{{\cal C}}

\def\CM{{\cal M}}
\def\CN{{\cal N}}

\def\CS{{\cal S}}
\def\CT{{\cal T}}
\def\cU{\mathcal{U}}
\def\cV{\mathcal{V}}

\def\fso{\mathfrak{so}}

\def\su{\mathfrak{su}}

\def\sp{\mathfrak{sp}}

\def\e{\mathfrak{e}}
\def\g{\mathfrak{g}}
\def\f{\mathfrak{f}}

\newcommand{\intpart}{_\text{int}}

\usepackage{xcolor}

\begin{document}

\title{6d \Nequals{(1,0)} theories on $S^1/T^2$ \\
and class S theories: part II}
\abstract{
We study the $T^2$ compactification of a class of 6d \Nequals{(1,0)}  theories
that is Higgsable to \Nequals{(2,0)} theories.
We show that the resulting 4d \Nequals{2} theory at the origin of the Coulomb branch and the parameter space is generically given by two superconformal matter sectors coupled by an infrared-free gauge multiplet and another conformal gauge multiplet. Our analysis utilizes the 5d theories obtained by putting the same class of 6d theories on $S^1$.

Our class includes, among others, the 6d theories describing multiple M5 branes on an ALE singularity, and we analyze them in detail. The resulting 4d theory has manifestly both the $\SL(2,\bZ)$ and the full flavor symmetry. We also discuss in detail the special cases of 6d theories where the infrared-free gauge multiplet is absent.

In an appendix, we give a field-theoretical argument for an F-theoretic constraint that forbids a particular 6d anomaly-free matter content, as an application of our analysis.
}
\author[1]{Kantaro Ohmori,}
\author[1]{Hiroyuki Shimizu,}
\author[1,2]{Yuji Tachikawa,}
\author[3]{and Kazuya Yonekura}
\affiliation[1]{Department of Physics, Faculty of Science, \\
 University of Tokyo,  Bunkyo-ku, Tokyo 133-0022, Japan}
\affiliation[2]{Institute for the Physics and Mathematics of the Universe, \\
 University of Tokyo,  Kashiwa, Chiba 277-8583, Japan}
\affiliation[3]{School of Natural Sciences, Institute for Advanced Study,\\
Princeton, NJ 08540, United States of America}
\preprint{IPMU-15-0017, UT-15-25}

\maketitle

\section{Introduction and summary}\label{sec:introduction}
In \cite{Ohmori:2015pua}, we started the analysis of the $T^2$ compactification of 6d \Nequals{(1,0)} theories.
There, we concentrated on a class of theories which we called \emph{very Higgsable}, namely those theories that have a Higgs branch where no tensor multiplet remains. 
We found, in that case, that the 4d \Nequals{2} theory at the origin of the Coulomb branch and the parameter space is naturally a superconformal field theory (SCFT), whose anomaly polynomial is given in terms of the anomaly polynomial of the parent 6d theory.
In the F-theoretic construction of 6d SCFTs of \cite{Heckman:2013pva,DelZotto:2014hpa,Heckman:2015bfa},
the very Higgsable theories correspond to the case where all of the compact cycles producing the tensor multiplets can be removed by repeated blow-downs of ($-1$)-curves.\footnote{Purely field-theoretically,
having a $(-n)$-curve means that i) there is a tensor multiplet, ii) whose scalar component gives the tension of a stringy excitation, iii) such that the Dirac charge quantization pairing of a string with itself is $n$.  In the rest of the paper, we use the F-theory language for convenience, but what we need is the existence of an ultraviolet-complete theory at the origin of the moduli space, and not the F-theory construction itself.
} This class includes the E-string theories of general rank and the 6d theories of a single M5-brane probing an ALE singularity. 

A natural next step in the analysis would be, then, to study the compactification of the class of 6d theories that have only ($-2$)-curves at the endpoint where all possible blow-downs of $(-1)$-curves are performed.
We call such a theory \emph{Higgsable to \Nequals{(2,0)} theories}.
This is because we can modify the complex structure moduli of the F-theory setup so that $(-2)$-curves do not have any decoration, meaning that we can go to a point on the Higgs branch where the low-energy theory is just the \Nequals{(2,0)} theory.
This class includes, among others, the 6d theories describing multiple M5-branes on an ALE singularity, called \emph{conformal matters} in \cite{DelZotto:2014hpa}.

\paragraph{Conventions:}
Before proceeding, we list some conventions to be used in this paper.
We reserve the capital letter $G$  for the type of the Dynkin diagram formed by the $(-2)$-curves, and the corresponding group.  
We typically do not distinguish groups sharing the same Lie algebra, unless necessary.
Quantum field theories are denoted by curly alphabets such as $\CT$ or $\CS$.
Class S theories are considered as known, and we reserve sans-serif letters for them; so the $T_N$ theory is denoted as $\sT_N$.
We also use the following notations as in  \cite{Tachikawa:2015bga}, to denote various operations on quantum field theories:
\begin{itemize}
\item the theory $\CT\vev{M}$ stands for the compactification on $M$ of a theory $\CT$, 
\item the notation $\CT\{H\}$ means that the theory $\CT$ has the flavor symmetry $H$, and
\item the theory $\CT/H$ stands for a gauge theory where  a $H$ gauge multiplet couples to $\CT\{H\}$ via its $H$ flavor symmetry.
\end{itemize}

\paragraph{The 6d theory:}
Take such an \Nequals{(1,0)} theory $\CT^\text{6d}$, Higgsable to the \Nequals{(2,0)} theory of type $G$, where $G=A_{n-1},$ $D_n$ or $E_n$. 
By definition of $\CT^\text{6d}$, there is a subspace of the tensor branch where only tensor multiplets associated to the underlying $\CN=(2,0)$ theory
get vacuum expectation values.
We call this branch as $\CCt$, and throughout the paper, we mean this $\CCt$ whenever we say tensor branch unless otherwise stated.
On the generic points on this tensor branch $\CCt$, $\CT^\text{6d}$ has
\begin{itemize}
\item  $\rank G$ tensor multiplets corresponding to $\rank G$ $(-2)$-curves,
\item simple gauge algebras $\fg_i$, $(i=1,\ldots,\rank G)$, whose couplings are specified by the vevs of the scalar components of the tensor multiplets above, and
\item various very Higgsable bifundamental matter SCFTs $\cH_{ij}^\text{6d}$ 
with flavor symmetry $\fg_i\times \fg_j$ when
the $i$-th and $j$-th $(-2)$-curves intersect,
and some additional very Higgsable  theory $\cH_i^\text{6d}$ charged under  a single algebra $\fg_i$.
\end{itemize}
Note that some of $\fg_i$ can be empty, which we often formally write as $\fg_i=\su(1)$.

As an example, the worldvolume theory on $n$ M5 branes on the $\bC^2/\bZ_k$ singularity, without the center-of-mass multiplet, can be Higgsed to $G=A_{n-1}$ \Nequals{(2,0)} theory. 
On the tensor branch, we have $\fg_{1,\ldots,n-1}=\su(k)$, 
in addition we have flavor symmetries $\fg_0=\fg_n=\su(k)$,
and there are bifundamental hypermultiplets of $\fg_i\times \fg_{i+1}$ for $i=0,\ldots,n-1$ as $\cH_i$ and $\cH_{ij}$. 
That is, we have
\begin{equation}
\CT^\text{6d} \xrightarrow{\text{tensor branch}}
\text{$\rank G$ tensors}+\text{6d generalized quiver theory}
\end{equation} 
where \begin{equation}
\text{6d generalized quiver theory}=
\frac{\prod_{i,j} \cH^\text{6d}_{ij}\{\fg_i,\fg_j\} \times \prod_i \cH^\text{6d}_i\{\fg_i\} }{\prod_i \fg_i}.
\end{equation}
Our assumption is that all the matter theories $\cH_{ij}$ and $\cH_i$ are very Higgsable,
and  strings associated to each $\fg_i$ have the Dirac quantization pairing 2 with themselves.

\paragraph{The 5d theory:}
Our interest is the $S^1$ or $T^2$ compactification of  $\CT^\text{6d}$, at the most singular point on its Coulomb branch. We claim and provide ample pieces of evidence to the following facts:

\begin{claim}
1. Consider the  theory $\CT^\text{5d}=\CT^\text{6d}\vev{S^1_R}$, namely the 5d theory obtained by compactifying the 6d theory  on an $S^1$ of radius $R$.
At the most singular point of the moduli and parameter space, this 5d theory is given by an \Nequals{1} vector multiplet of gauge group $G$,
together with  a 5d SCFT we denote as $\CS^\text{5d}\{G\}$ 
whose  $G$ flavor symmetry  is gauged by the vector multiplet. 
The gauge coupling of the vector multiplet is given by $8 \pi^2/g_G^2 = R^{-1}$. Using our notation, we just have \begin{equation}
\CT^\text{6d}\vev{S^1_R} =  \CT^\text{5d} = \CS^\text{5d}\{G\} / G_{8 \pi^2/g_G^2 = R^{-1}}.\label{eq1}
\end{equation}
The 5d theory $\CS^\text{5d}\{G\}$ is such that when it is given a generic mass deformation to the flavor symmetry $G$, we have 
\begin{equation}
\CS^\text{5d}\{G\} \xrightarrow{\text{deform by $m$}}
\text{5d quiver theory}
\end{equation}
where
\begin{equation}
\text{5d quiver theory}
=\frac{\prod_{i,j} \cH^\text{5d}_{ij}\{\fg_i,\fg_j\} \times \prod_i \cH^\text{5d}_i\{\fg_i\} }{\prod_i \fg_i}
\end{equation}
\end{claim}
We will find, in fact, that when all $\fg_i$ are $\su$ gauge groups and all $\cH_{ij}$ and $\cH_i$ are hypermultiplets, $\CS^\text{5d}\{G\}$ actually has $G\times G$ symmetry.
In that case, the $G$ flavor symmetry in the notation $\CS^\text{5d}\{G\}$ denotes the diagonal subgroup of the $G\times G$ symmetry.

Note that when $\CT^\text{6d}$ is the \Nequals{(2,0)} theory of type $G$, $\CS^\text{5d}\{G\}$ is a free hypermultiplet in the adjoint representation of $G$, and Eq.~\eqref{eq1} is the standard relation between the \Nequals{(2,0)} theory in 6d and the \Nequals2 super Yang-Mills theory in 5d. 

Note also that when $\CT^\text{6d}$ is the theory on $N$ M5 branes on the $\bC^2/\bZ_k$ singularity, $\CS^\text{5d}\{G\}$ is the strongly-coupled SCFT whose mass deformation is the linear quiver of the form $[\su(k)_L]\times\su(k)^{n-1}\times [\su(k)_R]$. 
It has the flavor symmetry $\su(k)_L\times \su(k)_R\times \su(n)_L\times \su(n)_R$, and the flavor symmetry $G$  is the diagonal subgroup of $\su(n)_L\times \su(n)_R$.

\paragraph{The 4d theory:}
Our basic claims in the $T^2$ compactification can be summarized as follows:
\begin{claim}
2. Consider the 4d theory $\CS^\text{4d}\{G\}$ obtained by compactifying the theory $\CS^\text{5d}\{G\}$ we obtained above, on another  $S^1$.
At the most singular point of the moduli and parameter space, the 4d theory has the following structure: $\CS^\text{4d}\{G\}$ is a combined system \begin{equation}
		\CS^\text{4d}\{G\} = \frac{\cU^\text{4d}\{G,H\}  \times  \cV^\text{4d}\{H\}}{H}
	\end{equation}
	where $\cU$, $\cV$  are two 4d \Nequals2 SCFTs with the specified flavor symmetry, for a certain infrared-free group $H$. When $H=\varnothing$ and $\cV\{H\}$ is trivial, 
	$
	 \CS^\text{4d}\{G\} = \cU\{G,\varnothing\}
	$
 is itself a 4d \Nequals{2} SCFT. This happens, for example, when some of $\fg_i$ is empty. The flavor symmetry central charge $k_G$ of $\CS^\text{4d}\{G\}$ is the same as that of a free hypermultiplet in the adjoint of $G$. 
	
Furthermore, when we perform a diagonalizable mass deformation $m$ for the $G$ flavor symmetry of $\CS^\text{4d}\{G\}$, we obtain a generalized quiver theory \begin{equation}
\CS^\text{4d}\{G\} \xrightarrow{\text{deform by $m$}}
\text{4d quiver theory}
\end{equation}
where
\begin{equation}
\text{4d quiver theory}=\frac{\prod_{i,j} \cH^\text{4d}_{ij}\{\fg_i,\fg_j\} \times \prod_i \cH^\text{4d}_i\{\fg_i\} }{\prod_i \fg_i}
\end{equation}
where $\cH^\text{4d}_{ij}$ and $\cH^\text{4d}_{i}$ is the 4d SCFT obtained by the $T^2$ compactification of the very Higgsable 6d theory $\cH^\text{6d}_{ij}$ and $\cH^\text{6d}_i$
 such that all the couplings of $\fg_i$ are exactly marginal. 
\end{claim}

We give an algorithm to determine the infrared-free gauge group $H$ when $G$ and $\fg_i$ are all of type $A$,
and when $G$ is type $A$, $\fg_i=\fg$  for all $i$ and all the generalized matters are minimal conformal matters. 
In the former case, both $\cU$ and $\cV$ are class S theories of type $A$.
In the latter case, $\cV$ is a class S theory of type $\fg$ but we have not been able to identify $\cU\{G,H\}$ in general. %We at least give the formula to find the central charges of $\cU\{G,H\}$.

For example, when $\CT^\text{6d}$ is the theory on $N$ M5 branes on the $\bC^2/\bZ_k$ singularity, the corresponding $\CS^\text{4d}\{G\}$ is a coupled system of a class S theory of type $A_{k-1}$ and another class S theory of type $A_{N-1}$.
The precise structure is given in \eqref{eq:4dS}. 

We will also have an extensive discussion of the structure of the 4d SCFT $\cU\{G,\varnothing\}$ when the infrared-free $H$ is absent. We will determine its conformal central charges $n_v$, $n_h$ and the flavor central charges $k$ in terms of the coefficients of the 6d anomaly polynomial. The final formulas are given in \eqref{eq:baseAformula}, \eqref{eq:baseDformula1} and \eqref{eq:baseDformula2}.

Given the structure of $\CS^\text{4d}\{G\}$, it is easy to state the structure of the $T^2$ compactification of the 6d theory itself:
\begin{claim}
3. The $T^2$ compactification $\CT^\text{6d}\vev{T^2_\tau}$  of $\CT^\text{6d}$,
where $\tau$ is the complex structure moduli of the torus, at the most singular point on its Coulomb branch, has the structure \begin{equation}
\CT^\text{6d}\vev{T^2_\tau} = \frac{\cU^\text{4d}\{G,H\}  \times  \cV^\text{4d}\{H\}}{G_\tau \times H}
\end{equation} where the complexified inverse squared coupling of the $G$ vector multiplet is given by $\tau$.
The coupling of $G$ is exactly marginal and shows the $\SL(2,\bZ)$ duality symmetry,
and the coupling of $H$ is infrared free when present. 

Furthermore, by giving a generic vev  to the scalar component of the $G$ vector multiplet, we have
\begin{equation}
\CT^\text{6d}\vev{T^2_\tau} \xrightarrow{\text{generic vev for $G$}} 
\U(1)^{\rank G}+ \text{4d quiver theory}
\end{equation} where the couplings of $\fg_i$ are all exactly marginal 
and are controlled by the vevs of the scalars of the $\U(1)^{\rank G}$ vector multiplet.
\end{claim}

\paragraph{Organization:}
The paper is organized as follows.
We start in Sec.~\ref{sec:simplest} by whetting the appetite of the reader, 
by analyzing the $T^2$ compactification of the 6d \Nequals{(1,0)} theory describing two M5-branes probing an ALE singularity.
Using the analysis as in \cite{Ohmori:2015pua},
we will show that the 4d system at the most singular point in the Coulomb branch is described by a class S theory coupled to the \Nequals4 $\SU(2)$ super Yang-Mills via an additional infrared free $\SU(2)$ gauge group.

In Sec.~\ref{sec:general}, we first recall the general structure of 6d \Nequals{(1,0)} theories of our interest. The non-renormalization theorem we prove in this section plays a significant role throughout the paper. Then we analyze and describe the general structure of the $S^1$ and $T^2$ reduction of our class of the 6d \Nequals{(1,0)} theories. 
When the 6d theory has ($-2$)-curves arranged in a Dynkin diagram of type $G$, we show that the 5d theory is a $G$-symmetric 5d \Nequals1 SCFT coupled to an infrared-free $G$ gauge multiplet. 
The 4d physics can then be understood by the $S^1$ reduction of this $G$-symmetric 5d SCFT.

In Sec.~\ref{sec:examples1} and Sec.~\ref{sec:examples2}, we flesh out the discussions in the previous section by studying various concrete examples, including and vastly extending the cases studied in Sec.~\ref{sec:simplest}. 
Specifically, we discuss the compactification of the conformal matters of \cite{DelZotto:2014hpa}. Sec.~\ref{sec:examples1} deals with the A-type conformal matters and Sec.~\ref{sec:examples2} deals with the general type conformal matters. Note that our general argument gives several new predictions: about T-duality between little string theories and about collisions of punctures in the class S theory of type $E_7$ and $E_8$.

In Sec.~\ref{sec:Coulomb}, we perform the detailed study of the subbranch of the 4d Coulomb branch for the 6d theory on $T^2$, where only Coulomb moduli coming from the tensor multiplets get vevs. In the way, we propose the generalization of Argyres-Seiberg-Gaiotto dualities \cite{Argyres:2007cn,Gaiotto:2009we}. We also obtain a sufficient condition for obtaining a 4d \Nequals{2} SCFT without IR free gauge groups at the most singular point of the 4d Coulomb branch. We relate the central charges of that 4d \Nequals{2} SCFT to the anomaly polynomial of the parent 6d \Nequals{(1,0)} theory, extending the analysis in \cite{Ohmori:2015pua}.

We have an appendix~\ref{appendix} where we give a field-theoretical argument why the configurations $\begin{array}[b]{cc}
\mathfrak{su}(2) & \mathfrak{so}(7)\\
2&2
\end{array}$ with the matter content $\frac12(\mathbf{2},\mathbf{7})$ 
and
$\begin{array}[b]{cc}
\mathfrak{su}(2) & \mathfrak{so}(8)\\
2&n
\end{array}$ with $n=1,2,3$
are not allowed, as an application of our analysis. These constraints were originally found F-theoretically, see e.g.~Sec.~6.2.1 of \cite{Heckman:2015bfa}.

Before proceeding, we note that there recently appeared a paper \cite{DelZotto:2015rca} where 
the $T^2$ compactification of many classes of 6d \Nequals{(1,0)} theories were studied in terms of their Seiberg-Witten solutions, 
and many of the theories we discuss have already been analyzed there. 
We believe our paper provides complementary information about them, as our approach does not directly use the stringy features of F-theory as they did, and our focus is about the most singular point in the Coulomb branch whereas they mainly considered less singular points.

\section{Two M5-branes probing an ALE singularity on $T^2$}
\label{sec:simplest}

\subsection{Two M5-branes on $\bC^2/\bZ_2$}
\label{sec:simplestX}
Let us start with one of the simplest theories of our class, namely the 6d theory describing two M5-branes on $\bC^2/\bZ_2$. 
Here and in the following, we always discard the center-of-mass mode.
Then this theory has one tensor branch direction corresponding to the separation of the two M5-branes along the singularity.
On a generic point of the tensor branch, we have an $\SU(2)$ gauge group coming from the $\bC^2/\bZ_2$ singularity, coupled to four hypermultiplets in the doublet, 
two coming from the M5-brane on the left, another two from the one on the right. 
The $\SU(2)$ coupling is given by the separation of the M5-branes, i.e.~the vacuum expectation value (vev) of the tensor multiplet scalar.

Compactify the theory on $T^2$ with modulus $\tau$. 
The Coulomb branch is complex two dimensional. 
One direction $u$ comes from the tensor branch vev, and is cylindrical at the asymptotic infinity since one real direction comes from 
integrating the two-form of the tensor multiplet on $T^2$.
On a generic point, there is a $\U(1)_u$ group associated to the scalar $u$.
Another direction $v$ is the vector multiplet scalar of the \Nequals2 $\SU(2)_v$ theory with four flavors.
Since this $\SU(2)_v$ theory is superconformal, there is a natural origin of the $v$ direction at each value of $u$.
This determines a natural one-dimensional subspace $\CCt$ of the Coulomb branch.
By a slight abuse of terminology, we call this subspace  the 4d tensor branch.
The $v$ direction is fibered over this 4d tensor branch.

To determine the structure of $\CCt$, we go to the Higgs branch of the $\SU(2)_v$ theory, 
that correspond to moving the M5-branes away from the $\bC^2/\bZ_2$ singularity.
We then just have two M5-branes on a flat space, and we know $\U(1)_u$ enhances to \Nequals 4 $\SU(2)_u$ when the two M5-branes become coincident. 

Using the standard fact about the decoupling of the Higgs branch direction and the Coulomb branch direction \cite{deWit:1984px,Argyres:1996eh},
we see that the 4d tensor branch of our theory has exactly the same structure with that of \Nequals{(2,0)} theory of type $\SU(2)$ put on $T^2$, or equivalently the 5d maximally supersymmetric Yang-Mills with the gauge group $\SU(2)$ on $S^1$.
We can therefore introduce the coordinate system to the 4d tensor branch so that the $u$ coordinate is a cylinder with  identifications \begin{equation}
u\sim u+1, \qquad u \sim -u\label{eq:simple4dtensorbranch},
\end{equation} 
where the first corresponds to shifting the holonomy around $S^1$ by $2\pi$,
and the second to the Weyl symmetry.
We see that there are singularities at $u=0,1/2$ and $u\to \pm \ii \infty$. 

The coupling $\tau_u$ of $\U(1)_u$ is given by the geometric modulus $\tau$ of $T^2$, i.e.~$\tau_u=\tau$, and is constant over $\CCt$. 
Let us next discuss the coupling $\tau_v$ of the $\SU(2)_v$ theory with four flavors.
This is now a holomorphic function $\tau_v(u)$ on this 4d tensor branch $\CCt$, and it becomes weakly coupled as $u\to\pm \ii\infty$.
As the $\SU(2)_v$ theory is a conformal theory whose space $\CM_\text{marginal}$ of the exactly marginal coupling constant is nontrivial, $\tau_v(u)$ is better thought of as a holomorphic map \begin{equation}
\pi: \CCt \to \CM_\text{marginal}.
\end{equation} 
Let us realize $\CM_\text{marginal}$ in the description introduced by Gaiotto \cite{Gaiotto:2009we}, as $z\in\CP^1$ minus three points $0$, $1$ and $\infty$ as always, so that going around $z=0$ once corresponds to shifting the theta angle by $2\pi$. 
The holomorphy then uniquely fixes the map to be \begin{equation}
z=\pi(u)=e^{2\pi\ii u}.
\end{equation} Note that the Weyl symmetry $u \sim -u$ is realized as the automorphism of $\CP^1$ 
as $\{0,z,1,\infty\} \to \{\infty, z^{-1}, 1 ,0\} $.
The coupling $\tau_v(u)$ can be reconstructed from this map $\pi$ using the standard procedure. 
For example, in the weak coupling limit $z \to 0$ it is given as $\tau_v(u)=\frac{\theta}{2\pi} + \frac{4\pi\ii}{g^2} \sim \frac{1}{ 2 \pi \ii } \log z$.

In particular, this means that close to $u=0$ or equivalently $z=1$, we need to go to an S-dual frame of $\SU(2)_v$ theory with four flavors to have a weakly-coupled description.
In this new duality frame, the coupling $\tau_{v,D}=\frac{\theta_D}{2\pi} + \frac{4\pi\ii}{g^2_D}$ of the dual $\SU(2)_{v}$ gauge multiplet behaves as   \begin{equation}
\tau_{v,D} \sim   \frac{1}{2\pi \ii }\log (z-1) \sim  \frac{1}{2\pi \ii }\log u.\label{eq:infra-running}
\end{equation} 
This clearly shows that $\SU(2)_v$ is now infrared-free at $u=0$, 
meaning that there should be new light degrees of freedom charged under $\SU(2)_v$ there.\footnote{The logic here is analogous to
that of Seiberg and Witten \cite{Seiberg:1994rs}, except the fact that we are discussing the gauge coupling of the non-abelian $\SU(2)_v$
gauge group instead of $\U(1)$. The role of massless monopoles of \cite{Seiberg:1994rs} is played by the
new light degrees of freedom charged under the dual of $\SU(2)_v$.}
This running is the same one with that of $\SU(2)_v$ with five flavors.

This behavior can be explained if we posit that $\U(1)_u$  is enhanced to \Nequals4 $\SU(2)_u$.
Regarded as an \Nequals2 theory, it has an $\SU(2)$ flavor symmetry, and at nonzero $u$, 
it gives rise to one doublet hypermultiplet with mass $u$.
This is exactly the right number of massive hypermultiplets to produce the running \eqref{eq:infra-running}
when this $\SU(2)$ flavor symmetry is identified with  the $\SU(2)_v$ gauge symmetry. 
It is this $\SU(2)_u$ that has the coupling $\tau$ determined geometrically by the modulus of the torus used in the compactification, on which $\SL(2,\bZ)$ acts naturally. 

At the end of the day, we find that the most singular point $u=0$ on the 4d tensor branch $\CCt$ is described by an  $\SU(2)_u\times \SU(2)_{v,\text{dual}}$ gauge theory with the matter content $\frac12(\mathbf{3},\mathbf{2}) + \frac72(\mathbf{1},\mathbf{2})$. 
In the class S notation, the 4d theory we identify is 
\begin{equation}
\begin{tikzpicture}[baseline=(C),yscale=.9,xscale=-.9]
\node[flavor] (A) at (0,1) {$2$};
\node[flavor] (B) at (0,-1) {$2$};
\node[T,shape border rotate=180] (T1) at (1,0) {$\sT_2$};
\node[gauge] (C) at (2.5,0) {$2_v$};
\node[flavor] (D) at (2.5,1) {$1$};
\node[T] (T2) at (4,0) {$\sT_2$};
\node[gauge] (E) at (5,0) {$2_u$};
\draw (A)--(T1.right corner);
\draw (T1.left corner)--(B);
\draw (T1)--(C)--(T2);
\draw (C)--(D);
\draw[out=60,in=90] (T2.left corner) to (E.north);
\draw[out=-60,in=-90] (T2.right corner) to (E.south);
\end{tikzpicture}.\label{eq:tikz}
\end{equation}
Note that when the $\SU(2)_u$ is broken to $\U(1)_u$ on the Coulomb branch, this correctly reduces to $\SU(2)_v$ theory with four massless flavors and one massive flavor.
If we keep only the massless ones, this is the S-dual of the $\SU(2)_v$ with four flavors that descends from the 6d.
This structure will be used in Appendix~\ref{appendix}.

\subsection{Two M5-branes on other ALE singularities}
It is straightforward to generalize the discussion to the case of two M5-branes probing an ALE singularity of more general type.

First, consider the case with $\bC^2/\bZ_k$.
We still have a 4d tensor branch $\CCt$ described by \eqref{eq:simple4dtensorbranch} with coordinate $u$.
When the value of $u$ is generic, we have 4d \Nequals2 $\SU(k)_{\vec v}$ theory with $2k$ flavors,
providing $k-1$ additional directions in the Coulomb branch we collectively denote by $\vec v$.
The coupling  $\tau_v$ of $\SU(k)_{\vec v}$ is still given as before.
Close to the singularity $u=0$, we need to go to the S-dual frame, whose description is by now familiar \cite{Gaiotto:2009we,Chacaltana:2010ks}.
Here we use the notation of \cite{Tachikawa:2015bga}.
We have a $\sT_k\{[1^k],[1^k],[k-2,1^2]\}$ whose $\SU(2)$ flavor symmetry of the puncture of type $[k-2,1^2]$ is gauged by $\SU(2)_v$ with an additional doublet.
The coupling of this conformal $\SU(2)_v$ is given by $\tau_D$ in \eqref{eq:infra-running}.
This running of $\tau_{v,D}$ can be physically accounted for, 
again by the enhancement of $\U(1)_u$ to $\SU(2)_u$,
whose coupling $\tau_u$ is given geometrically by the modulus of the torus $T^2$ used in the compactification.

The 4d theory at the origin of the 4d tensor branch is thus described in the class S language as
\begin{equation}
\begin{tikzpicture}[baseline=(C),yscale=.9,xscale=-.9]
\node[flavor] (A) at (0,1) {$k$};
\node[flavor] (B) at (0,-1) {$k$};
\node[T,shape border rotate=180] (T1) at (1,0) {$\sT_k$};
\node[gauge] (C) at (3.5,0) {$2_v$};
\node[T] (T2) at (5,0) {$\sT_2$ };
\node[gauge] (E) at (6,0) {$2_u$};
\draw (A)--(T1.right corner);
\draw (T1.left corner)--(B);
\draw (T1) to node[pos=0.1,above]{$\scriptstyle [k-2,1^2]$} (C)--(T2);
\draw[out=60,in=90,] (T2.left corner) to (E.north);
\draw[out=-60,in=-90] (T2.right corner) to (E.south);
\end{tikzpicture}
\end{equation}
where the decoration at the third vertex of the left triangle denotes the type of the third puncture.
Note that the $\SU(2)$ symmetry central charge of the puncture $[k-2,1^2]$ is equal to that of three flavors,
and therefore no need to add any additional doublet to reproduce the running \eqref{eq:infra-running}.

Second, let us consider the case when the singularity is $\bC^2/\Gamma_{D_k}$.
Let us concentrate on the one-dimensional subspace of the 6d tensor branch corresponding to separating two M5-branes without fractionating each of the full M5-branes. 
On this tensor branch, we have the 6d gauge group $\fso(2k)$, and each of the M5-branes on the left and on the right give a single conformal matter of type $\fso(2k)$, which is a strongly-coupled SCFT. 
Reducing on $T^2$, we still have a 4d tensor branch with the coordinate $u$ as before,
and on its generic point, the gauge group is coupled to two copies of class S theory of type $\fso(2k)$ with two full punctures and one simple puncture \cite{Ohmori:2015pua}.
Stated differently, we just have a class S theory of type $\fso(2k)$ with two full punctures and two simple punctures. 
The structure of the 4d tensor branch is entirely analogous to the cases treated above.
Close to $u=0$, we need to go to the S-dual frame.
The simple puncture is of type $[2k-3,3]$, and colliding two of them, we get a puncture of type $[2k-5,2^2,1]$ whose $\SU(2)$ flavor symmetry central charge is equal to that of seven half-hypermultiplets in the doublet \cite{Chacaltana:2011ze}. 
Therefore we need one additional doublet half-hypermultiplet to make the dual $\SU(2)$ gauge group superconformal. 
To account for the running \eqref{eq:infra-running}, we again expect that $\U(1)_u$ to enhance to $\SU(2)_u$.
The resulting 4d theory then has the form \begin{equation}
\begin{tikzpicture}[baseline=(C),xscale=-1]
\node[flavor] (A) at (-1,1) {$\fso(2k)$};
\node[flavor] (B) at (-1,-1) {$\fso(2k)$};
\node[T,shape border rotate=180] (T1) at (1,0) {$\sT_{D_k}$};
\node[gauge] (C) at (3.5,0) {$2_v$};
\node[gauge] (E) at (5.5,0) {$2_u$};
\draw (A)--(T1.right corner);
\draw (T1.left corner)--(B);
\draw (T1) to node[pos=0.1,above]{$\scriptstyle [2k-5,2^2,1]$} (C)--(T2);
\draw (C) to node[pos=0.1,above]{$\mathbf{2}$} node[pos=0.5,below]{half} node[pos=0.9,above]{$\mathbf{3}$} (E);
\end{tikzpicture}
\end{equation}
where the matter content charged under $\SU(2)_v\times \SU(2)_u$ is the half hypermultiplet in $(\mathbf{2},\mathbf{3})$.

The analysis of the case with $\bC^2/\Gamma_{E_6}$ is entirely analogous.  Using the tables in \cite{Chacaltana:2014jba}, we see that the 4d theory is given by 
\begin{equation}
\begin{tikzpicture}[baseline=(C),xscale=-1]
\node[flavor] (A) at (-1,1) {$E_6$};
\node[flavor] (B) at (-1,-1) {$E_6$};
\node[T,shape border rotate=180] (T1) at (1,0) {$\sT_{E_6}$};
\node[gauge] (C) at (3.5,0) {$2_v$};
\node[gauge] (E) at (5.5,0) {$2_u$};
\draw (A)--(T1.right corner);
\draw (T1.left corner)--(B);
\draw (T1) to node[pos=0.1,above]{$\scriptstyle A_5$} (C)--(T2);
\draw (C) to node[pos=0.1,above]{$\mathbf{2}$} node[pos=0.5,below]{half} node[pos=0.9,above]{$\mathbf{3}$} (E);
\end{tikzpicture}.
\end{equation}
The remaining case with $E_7$ and $E_8$ cannot yet be studied as the analysis of the  collision of two simple singularities there has not been published. 
Judging from the tables in \cite{Chacaltana:2012zy}, it is plausible that it is entirely analogous, save the  type of the third puncture in the class S theory of type $E_{7,8}$. 
Presumably they have the Bala-Carter label $E_6$ and $E_7$, respectively.

\section{General structure of theories on $(\mathbf{-2})$-curves}\label{sec:general}
In this section, we explain the structure of 6d \Nequals{(1,0)} theories we want to compactify and give general arguments for the $S^1/T^2$ compactification of these theories. The results in this section will be checked using several examples in the following sections.

\subsection{A brief review of structure of 6d SCFTs}
%\label{sec:review}
Let us first very briefly review the structure of 6d SCFTs constructed in F-theory \cite{Heckman:2013pva,DelZotto:2014hpa,Heckman:2015bfa}
to explain some terminology used later in the paper.
6d SCFTs can be constructed by F-theory on elliptic Calabi-Yau threefolds $CY_3$. The F-theory is on the space
${\mathbb R}^{1,5} \times CY_3$ with flat six dimensional space ${\mathbb R}^{1,5} $. 
The base $B_2$ of the elliptic fibration is a non-compact, complex 2 dimensional space.
In the base $B_2$, there are 2-cycles (complex curves with the topology $\mathbb{CP}^1$), $C^i$, which are intersecting with each other. 
The size of the curves is determined by vevs of \Nequals{(1,0)} tensor multiplets.
We consider a configuration of curves such that the negative of the intersection matrix, $\eta^{ij}=- C^i \cdot C^j$, is positive definite.
Then we can shrink all the curves $C^i$ simultaneously to zero size to get a singularity. This corresponds to taking the vevs of the tensor multiplets to be zero.
The 6d SCFTs are realized on this singularity.

If some of the curves have self-intersection $-1$, i.e., $C^i \cdot C^i=-1$, we can blow-down these curves without making
the base $B_2$ singular (but the elliptic fibration becomes singular). By successive blow-down of $-1$ curves, we reach a configuration of curves
in which the self-intersection of all the curves satisfy $C^i \cdot C^i \leq -2$. Such a configuration is called the endpoint.
Field theoretically, this is a subspace of the tensor moduli space of vacua.
One of the important properties of the endpoint is that it specifies the \emph{non-Higgsable} property of the theory~\cite{Morrison:2012np,Heckman:2013pva}.
Field theoretically, this means that no matter how we try to make the theory higgsed at a generic point of tensor branch, there still remain
tensor multiplets (and minimal gauge groups on them required by elliptic Calabi-Yau condition) which remain un-higgsed.

What was found in \cite{Heckman:2013pva,DelZotto:2014hpa,Heckman:2015bfa} is that 
we basically get a quiver gauge theory on the endpoint configuration.
Each curve $C^i$ supports a simple gauge group (which could be empty),
and there are ``generalized bifundamental matters'' between curves which are intersecting with each other and ``generalized fundamental matters'' for each curves.
These generalized matters are sometimes just hypermultiplets, but they can also be strongly interacting SCFTs.
In the above language, these generalized matters are SCFTs which are obtained from configuration of curves
whose endpoint is trivial. That is, there is no singularity in the base after the blow-down of curves of self-intersection $-1$, 
and only the fibers are singular. Such theories are called very Higgsable in \cite{Ohmori:2015pua} because 
it has a Higgs branch without any tensor or vector multiplets which corresponds to deforming elliptic fibration non-singular. 

Therefore, in the endpoint configuration, we get a quiver theory with very Higgsable generalized matters.
There are constraints on allowed endpoint configurations and gauge groups.
Among them, a class of allowed endpoint configurations is the case where the endpoint only contains ($-2$)-curves
which are intersecting according to the Dynkin diagram of a simply laced gauge group $G=A,D,E$.
When the elliptic fibration is non-singular at all, the theory is effectively type IIB string theory on $B_2$ 
and we get the \Nequals{(2,0)} theory of type $G$ \cite{Witten:1995zh}.
By making the fibration singular, we can get more general theories. These are the class of theories we want to discuss in this paper.

\subsection{Non-Higgsable component and nonrenormalization}

If we go to the Higgs branch of the theory as far as possible, we get a non-Higgsable theory which is the \Nequals{(2,0)} theory of the type $G$.
The Higgs branch is the same in any dimensions, and Higgs moduli fields and tensor/Coulomb moduli fields do not mix with each other in the effective action. 
We can consider a subspace $\CCt$ of the tensor/Coulomb
moduli space where only the moduli which originate from the tensor multiplets of the 6d theory get vevs.\footnote{
Since the 6d theory has the Higgs branch on which the theory flows to the \Nequals{(2,0)} theory along $\CCt$,
there is also a subspace of the 5d/4d Coulomb branch where the corresponding branch opens. This clearly defines the subspace $\CCt$ in 5d/4d.
}
Then, the effective action (or more specifically the kinetic terms) of moduli fields parameterizing $\CCt$  in 6d/5d/4d
is the same as that of the \Nequals{(2,0)} theory in 6d/5d/4d because these two theories are
smoothly connected by Higgs deformation which does not affect the tensor/Coulomb effective action.

The difference between the general theory we are considering and the \Nequals{(2,0)} theory is that 
the general theory contains more massless degrees of freedom other than the moduli fields of $\CCt$. However, we emphasize again that
the effective action of $\CCt$ moduli fields and in particular the position of the singular loci on $\CCt$ are the same as in the $\CN=(2,0)$ theory.
In other words, the moduli fields of $\CCt$ are not renormalized by the existence of additional massless degrees of freedom.
Due to \Nequals{(2,0)} supersymmetry of the Higgsed theory, they are not renormalized at all.

\subsection{$S^1$ compactification to five dimensions}
\label{sec:five}
Let us fix a 6d theory $\CT^\text{6d}$ that can be Higgsed to a \Nequals{(2,0)} theory of type $G$,
and consider its $S^1$ compactification.
We go to the origin of the moduli space of the 6d theory at which we get the 6d SCFT,
and compactify it on a circle with radius $R$. We do not include any Wilson lines on $S^1$ which correspond to mass deformations in 5d.
In this setup, our conjecture is the following:

\begin{claim}
The 5d theory $\CT^\text{5d}$ obtained by the $S^1$ compactification at the most singular point of the moduli and parameter space is given by an \Nequals{1} vector multiplet of gauge group $G$ 
which is coupled to a 5d SCFT we denote as $\CS^\text{5d} \{G \}$,
whose  $G$ symmetry is gauged by the vector multiplet. 
The gauge coupling of the vector multiplet is given by $8 \pi^2/g_G^2 = R^{-1}$.
\end{claim}

Here, we used  the notation introduced in \cite{Tachikawa:2015bga}, where the groups listed inside $\{\cdots\}$ are the flavor symmetries, and 
 our normalization of the gauge coupling is such that $8 \pi^2/g_G^2$ is the one-instanton action.
We also note here that, when all $\fg_i$ are $\su$ gauge groups and all $\cH_{ij}$ and $\cH_i$ are hypermultiplets, $\CS^\text{5d}\{G\}$ actually has $G\times G$ symmetry.
In that case, the $G$ flavor symmetry in the notation $\CS^\text{5d}\{G\}$ denotes the diagonal subgroup of the $G\times G$ symmetry.

The main reason behind this conjecture is the following. In 6d, we can higgs the theory to obtain the \Nequals{(2,0)} theory of type $G$.
If we compactify it on this Higgs branch, we get \Nequals{2} super Yang-Mills in 5d with gauge group $G$, and in particular, we get a vector multiplet
with gauge coupling $8 \pi^2/g_G^2 = R^{-1}$. Now we slowly turn off the Higgs vev.
The important point is that the Higgs moduli and Coulomb moduli do not mix with each other.
Then the existence of the vector multiplet with the gauge coupling $8 \pi^2/g_G^2 = R^{-1}$ does not change
in the process of turning off the Higgs vev, and hence the vector multiplet exists even at the origin of the moduli space.
This establishes the fact that the vector multiplet with gauge group $G$ and gauge coupling $8 \pi^2/g_G^2 = R^{-1}$ exists
in the 5d theory after compactification of the 6d SCFT. 

The existence of the vector multiplet can be regarded as a kind of no-go theorem; the 5d theory
cannot be completely superconformal, because we always have the IR free vector multiplet. Our conjecture is that
this vector multiplet is the only non-SCFT component in 5d, and the rest of the theory is really an SCFT which we denoted as $\CS^\text{5d} \{G \}$.
When $G$ is trivial, that is, when there are no ($-2$)-curves in the endpoint, the 6d theory is very Higgsable.
In this case, our conjecture above says that the 5d theory obtained by $S^1$ compactification of a 6d very Higgsable theory
is really a 5d SCFT. This statement has been indeed established in \cite{Ohmori:2015pua}.\footnote{There, it was shown that the $T^2$ compactification of very Higgsable theory is a 4d SCFT,  and the structure of the singularities on its Coulomb branch was also completely fixed. 
Taking the limit of very thin $T^2$, we can obtain the singularity structure of the Coulomb branch of the 5d theory, which shows that the origin of the 5d theory is superconformal.}

In the case of the \Nequals{(2,0)} theory, our 5d SCFT is just a hypermultiplet in the adjoint representation of $G$.
The story of the general case is quite similar to the case of the \Nequals{(2,0)} theory by replacing
the adjoint hypermultiplet with $\CS^{5d} \{G \}$. For example, instantons of the $G$ vector field
is expected to correspond to the Kaluza-Klein modes of the $S^1$ compactification as in \cite{Douglas:2010iu,Lambert:2010iw}.

\paragraph{Tensor branch effective action.}
We want to discuss some of the consequences of our conjecture. Before doing that, we need some preparation. 
Let $\eta^{ij}$ be the negative of the intersection matrix of the ($-2$)-curves. It is also the same as the Cartan matrix of $G$.
The bosonic components of the tensor multiplets are denoted as $(\phi_i, B_i)$, where $\phi_i$ are real scalars and $B_i$
are 2-forms whose field strengths are self-dual. The $B_i$ are normalized in such a way that their field strengths are in integer cohomology.

We raise and lower the indices by $\eta^{ij}$ and its inverse matrix.
The volume of the ($-2$)-curve labelled by $i$ is proportional to $\phi^i = \eta^{ij} \phi_j$, and hence the inverse gauge coupling squared
of the gauge field at the ($-2$)-curve is proportional to $\phi^i$. We denote the gauge field strength at the node $i$
as $F_i$, and normalize it in such a way that the factor of $1/ 2\pi$ is absorbed in $F_i$, e.g. $ i F_i$ is in integer cohomology if the group is $\U(1)$.
(See \cite{Ohmori:2014kda} for more details of our notation and conventions.)
A part of the effective action is given by
\begin{multline}
2 \pi  \int  \eta^{ij} \Bigl(-\frac{1}{2}d \phi_i \wedge \star d \phi_j- \frac{1}{2}dB_i \wedge \star dB_j \\
+ \phi_i (\frac{1}{4}\Tr F_j \wedge \star F_j)+ 
  B_i  (\frac{1}{4}\Tr F_j \wedge F_j ) \Bigr)
\end{multline}
where $\Tr $ is normalized in such a way that $\frac{1}{4} \Tr F^2$ gives 1 for one-instanton.
Here the action of $B_i$ is somewhat formal because its field strength is self-dual. But the action after dimensional reduction will have definite meaning.
The part containing the 2-form $B_i$ is required by Green-Schwarz anomaly cancellation, and the part containing $\phi_i$ is related to the $B_i$ part by 
supersymmetry.

After dimensional reduction to 5d, we define $\Phi_i = 2 \pi R \phi_i$ and $A_{i,\mu}=2 \pi R B_{i,\mu 5}$ and obtain
\begin{multline}
\int  \eta^{ij} \Bigl(-\frac{1}{2R} (d \Phi_i \wedge \star d \Phi_j+dA_i \wedge \star dA_j )\\
+ 
2\pi \Phi_i (\frac{1}{4}\Tr F_j \wedge \star F_j)+   2 \pi A_i  (\frac{1}{4}\Tr F_j \wedge F_j ) \Bigr).
\end{multline}
The configuration of ($-2$)-curves defines a Dynkin diagram. Let $H^i$ be the Cartan of the $\SU(2)$ subalgebra of the node $i$
normalized as $\tr (H^i H^j) = \eta^{ij}$, where $\tr$ is normalized in such a way that it coincides with the trace in the fundamental representation in
$\SU(2)$ subalgebras. 
Then $\Phi_i$ and $A_i$ can be identified as the Cartan part of the vector multiplet of the 
5d gauge group $G$ as $\Phi_G = 2\pi H^i  \Phi_i$ and $A_G= 2 \pi  H^i A_i $; here we restored the $2 \pi$ factors of $\Phi_G$ and $A_G$.
But the normalization of $F_i$ is still different from the usual one by $1/2\pi$.
Then the above action can be rewritten as 
\begin{multline}
 \int \Big(  -\frac{1}{g_G^2} \tr(d \Phi_G \wedge \star d \Phi_G + F_G \wedge \star F_G )  \\
+ \tr (H^j\Phi_G) (\frac{1}{4}\Tr F_j \wedge \star F_j) +\tr (H^j A_G)  (\frac{1}{4}\Tr F_j \wedge F_j ) \Big),  \label{eq:eff5d}
\end{multline}
where $8 \pi^2/g_G^2 = R^{-1}$. The first two terms are the action of the vector multiplet for the gauge group $G$ (on the Coulomb branch),
while the last two terms are the action of the gauge groups supported on the ($-2$)-curves.

\paragraph{Mass deformation of 5d SCFT and 5d quiver.}
Now let us see the implication of our conjecture. In 6d tensor branch, we have a quiver gauge theory
whose gauge groups are supported on the ($-2$)-curves. Bifundamentals and fundamentals are generalized matters which are very Higgsable.
If we compactify this tensor branch theory to 5d, we get the same quiver theory in 5d. 
The gauge couplings are determined by the vev of $\Phi_G$ as in \eqref{eq:eff5d}.
The bifundamentals and fundamentals are 5d version of the very Higgsable theories.

On the other hand, we conjectured that the 5d theory at the origin of the moduli space is a system in which a 5d SCFT $\CS^{5d}\{G\}$ is coupled to
the $G$ gauge field. Going to the tensor branch in 6d corresponds to giving vevs to the adjoint scalar $\Phi_G$ of the vector multiplet.
The adjoint vev gives mass deformation of this 5d SCFT $\CS^\text{5d}\{G\}$.
Therefore, our conjecture requires that the mass deformation of the $\CS^\text{5d}\{G\}$ flows under RG flow to the 5d quiver,
\beq
\CS^\text{5d}\{G\} \xrightarrow{\text{mass deformation}} \text{the 5d quiver theory },
\eeq
where the quiver theory is the one obtained from the 6d tensor branch.
Furthermore, \eqref{eq:eff5d} tells us that the gauge coupling of the gauge field at the quiver node $i$ is given
by the mass deformation $\vev{\Phi_G} =m_G$ as
\beq
\frac{8\pi^2}{g^2_i}= \tr (H^i m_G),\label{eq:quivercoup}
\eeq
where we have used the fact that our normalization is such that $\frac{1}{4} \Tr F^2$ is 1 in one-instanton.

Let us state the above process in the opposite direction of RG flows.
Our conjecture requires that the 5d quiver gauge theory must have a UV fixed point.
Furthermore, there must be an enhanced global $G$ symmetry in the UV fixed point whose Cartan part is identified with the topological $\U(1)$ symmetries associated to instantons of gauge groups in the IR quiver.
If all the matters of the IR quiver are hypermultiplets, which requires all the gauge groups are $\su$, the UV fixed point should in fact have the $G\times G$ symmetry whose Cartan part is the $\U(1)$ symmetries that act on matter hypermultiplets in the IR quiver, combined with the $\U(1)$ symmetries associated to instantons.
In that case, the gauging in $\CT^\text{5d}=\CS^\text{5d}\{G\}/G$ should be taken for the diagonal of the $G\times G$ symmetry so that
there is no commutant of the gauge group $G$ inside $G\times G$, because the $\U(1)$ symmetries which act on hypermultiplets are anomalous in 6d and hence should be absent in $\CT^\text{5d}$. 

Let us focus our attention to the case in which the gauge groups on the ($-2$)-curve of the node $i$ is $\su(N_i)$,\footnote{We use
the symbols $\su$ etc. for $\SU$ etc. gauge groups supported on ($-2$)-curves.}
where the rank $N_i$ can take arbitrary values as long as anomaly cancellation condition is satisfied.
Moreover we assume that all the matters in the quiver are just hypermultiplets and we do not have any strongly interacting generalized matters.\footnote{In most cases where all $\g_i$ are $\su$, the matters cannot be strongly coupled. Exceptions are $N=1$, $\g_1=\su(3)$ which couples to an E-string theory, and $N=2$, $\g_1=\g_2=\su(3)$ where the diagonal of two $\su(3)$ couples to an E-string theory.
}
Numerous studies of 5d $\su(N)$ quiver gauge theories have been done in the literature, see e.g.~\cite{Aharony:1997ju,Aharony:1997bh} and the references that cite them.
In the class of theories relevant to us, anomaly cancellation in 6d requires that all the matters are in the (bi)fundamental representations, and the total 
flavor number $N_{f, i}$ of the gauge group $\su(N_i)$ at each node of the quiver is given by $N_{f,i}=2N_i$. Also, there are no Chern-Simons
terms of the 5d $\su(N_i)$ gauge groups, since they come from dimensional reduction of 6d gauge groups.

In this case, the corresponding 5d quiver theory is expected to have a UV fixed point. The enhanced global symmetry in the UV fixed point is
actually two copies of $G$ \cite{Tachikawa:2015mha,Yonekura:2015ksa}, which we denote as $G_L \times G_R$.
We can take the diagonal subgroup $G_{\rm diag} \subset G_L \times G_R$, and deform the UV SCFT by
mass deformation of $G_{\rm diag}$ by $m_G$. 
Then the IR gauge coupling of the quiver is really given by the equation \eqref{eq:quivercoup}\footnote{See the last equation in 
section~3.4 of \cite{Yonekura:2015ksa}. The $\mathfrak{m}_\pm$ in that paper is taken to be $m_G$ here, and $H_i$ there is $\frac{1}{2}H^i$ here.}
Therefore, our conjecture works very well in this class of theories. 

More general case involves strongly interacting generalized matters.
Then it is not straightforward to study their 5d quivers. Nevertheless, we will discuss examples in the Sec.~\ref{sec:examples2}
in which such a quiver theory with generalized bifundamentals is dual to more conventional 
$\SU(N)$ quiver gauge theories with ordinary hypermultiplets. Existence of such examples supports our general conjecture.

\subsection{$T^2$ compactification to four dimensions}\label{3.4}
Let us denote by $\CS^{4d}\{ G \}$ the theory which is obtained by the $S^1$ compactification of the 5d SCFT $\CS^{5d}\{ G \}$.
This 4d theory $\CS^{4d}\{ G \}$ may be an SCFT or may contain IR free gauge groups; we will discuss this point in detail later in this paper.
Then, by compactifying the 5d theory of the previous subsection further on $S^1$, we get a theory in which 
the 4d vector multiplet of the gauge group $G$ is coupled to $\CS^{4d}\{ G \}$.
This is the theory we obtain by $T^2$ compactification. 
Therefore, the problem of $T^2$ compactification of the 6d SCFT is reduced to the problem of $S^1$ compactification of the 5d SCFT $\CS^{5d}\{ G \}$.

Let us determine the gauge coupling of the $G$ gauge field. For this purpose, we again use the reasoning of the previous subsections.
We can higgs the theory to obtain \Nequals4 super Yang-Mills in 4d. The Higgs and Coulomb moduli do not mix, so the higgsing does not affect
the gauge coupling of the $G$ gauge field. The gauge field of \Nequals4  super Yang-Mills is conformal with the gauge coupling given by the 
complex modulus $\tau$ of the $T^2$. Therefore, the $G$ gauge group before higgsing must also be conformal (i.e.,
has vanishing beta function) with the gauge coupling $\tau$.
The $\SL(2,{\bZ})$ of the $T^2$ acts on $\tau$, so the 4d theory has a nontrivial $\SL(2,{\bZ})$ S-duality group.
The fact that $G$ gauge group is conformal means that the theory $\CS^{4d}\{ G \}$ contributes to the beta function by the same
amount as that of one adjoint hypermultiplet.

\paragraph{Quiver on the tensor branch.}
By going to the tensor branch in 6d and compactifying it on $T^2$, or equivalently by giving a vev to the adjoint scalar of the $G$ vector multiplet
and mass-deforming $\CS^{4d}\{ G \}$ by that vev, we get a quiver gauge theory with generalized matters. The Cartan of the $G$ gauge field
becomes $\U(1)^{\rank G}$ free vector fields.

We now show that gauge groups of the quiver are conformal. For this purpose,
it is enough to concentrate on a single ($-2$)-curve. 
A little more generally, let $\g$ be a gauge group supported on 
a curve of self-intersection $-n$. The generalized matters coupled to this gauge group is very Higgsable,
and we denote the 6d anomaly polynomial of this very Higgsable theory as $I^{\rm vH}$.
Then the part of the anomaly polynomial of the total system containing the field strength of $\g$ is given as
\beq
I^{\rm vH}+I^{\rm vec}+I^{\rm GS},
\eeq 
where $I^{\rm vec}$ is the contribution from the $\g$ vector multiplet and $I^{\rm GS}$ is the Green-Schwarz contribution.
They contain \cite{Sadov:1996zm,Ohmori:2014kda}
\beq
I^{\rm vec} &\supset -\frac{h^\vee_\g}{12}p_1(T)c_2(\g), \\
I^{\rm GS}  &\supset  \frac{1}{2 n} ( \frac{2-n}{4}p_1(T)-nc_2(\g))^2 \supset  - \frac{2-n}{4}p_1(T) c_2(\g),
\eeq
where $p_1(T)$ is the first Pontryagin class of background metric, $c_2(\g) = \frac{1}{4}\Tr F^2$ is the second Chern class of $\g$
normalized so that one-instanton gives 1, and $h^\vee_\g$ is the dual Coxeter number of $\g$.
The terms containing $c_2(\g)$ must be cancelled in the total anomaly, so we get
\beq
I^{\rm vH} \supset \frac{1}{48} ( 4h^\vee_\g+12(2-n))p_1(T) c_2(\g).
\eeq
In \cite{Ohmori:2015pua}, it was shown that the coefficient of $p_1(T)c_2(\g)$ in the 6d anomaly polynomial of a very Higgsable theory is
proportional to the $\g$ flavor central charge of the corresponding 4d theory.
From the above result, it is given as $k_\g= 4h^\vee_\g+12(2-n)$. 
This $k_\g$ is the contribution of the very Higgsable theory to the 4d beta function of the $\g$ gauge group,
in the normalization that the vector multiplet contribution is $-4h^\vee_\g$. Therefore the beta function of $\g$ is proportional to 
$k_\g-4h^\vee_\g=12(2-n)$

From this we find the following fact: 
pick a $(-n)$-curve, supporting a gauge multiplet $\fg$ which is coupled to very Higgsable matters. 
In the 4d theory obtained by the $T^2$ reduction,  this gauge multiplet is  
\begin{itemize}
\item IR free when $n<2$,
\item conformal when $n=2$, and
\item asymptotic free when $n>2$.
\end{itemize}
In particular, in our theory with only ($-2$)-curves, the gauge groups are all conformal.

The gauge couplings of these conformal gauge groups are determined by the vev of the adjoint scalar $\Phi_G$.
When this vev is turned off, we get a more singular theory $\CS^{4d}\{ G \}$ coupled to the non-abelian $G$ group.
We stress that the flow from $\CS^{4d}\{ G \}$ to the quiver is mass deformation rather than exactly marginal deformation,
and hence some of the information is lost in the quiver theory because massive degrees of freedom are integrated out.
We have already seen examples of these phenomena in Sec.~\ref{sec:simplest}. More examples are given in Sec.~\ref{sec:examples1} and \ref{sec:examples2},
and general argument will be given in Sec.~\ref{sec:Coulomb}

\section{Conformal matters and class S theories, type $A$}
\label{sec:examples1}

In this section and the next, we give concrete examples of the general discussions of the previous section.
We focus on conformal matters \cite{DelZotto:2014hpa} and their deformation.

\subsection{Conformal matter of A-type }\label{sec:Aconf}
In M-theory, conformal matters are realized by $N$ M5-branes which are put on the singularities of ALE space of type $\g$.
Its tensor branch, in F-theory, is given by
\beq
\begin{array}{ccccccccc}
[\g] & \g & \cdots & \g &[ \g] \\
&  2&\cdots&2& 
\end{array}
\eeq
where there are $N-1$ curves of self-intersection $-2$ each of which has the gauge group $\g$, 
and bifundamentals between adjacent $\g$'s are minimal conformal matters (i.e., the theory with $N=1$) which are very Higgsable. 
The $\g$'s at the two ends are flavor symmetries which we denote as $\g_L$ and $\g_R$, respectively.
Note that the group $G$ of our discussion about conformal matters is always $G=\SU(N)$, since the configuration of the ($-2$)-curves is of $A_{N-1}$ type.
(The groups $G$ and $\g$ should not be confused.)

If we compactify the theory on $S^1$ with generic Wilson lines in the diagonal subgroup of the flavor symmetry $\g_L \times \g_R$,
we get a type IIA theory with $N$ D4-branes put on the singularity with generic $B$-flux through the singularity.
Then we get a quiver gauge theory \cite{Douglas:1996sw} 
whose nodes form an affine Dynkin diagram of type $\widehat{\g}$ and each node $k$ of the affine Dynkin diagram has
the gauge group $\SU(d_k^\g N)$, where $d_k^\g$ are the so-called marks of the Dynkin diagram 
such that the highest root is given by $\sum_k d_k^\g \alpha_k$ where $\alpha_k$ is the $k$-th simple root.
However, our main focus in this paper is to study the most singular theory obtained without flavor Wilson lines.

In this section we first consider  A-type conformal matters in which $\g=\su(k)$,
\beq
\begin{array}{ccccccccc}
[\su(k)_L ] & \su(k) & \cdots & \su(k) & [ \su(k)_R ] \\
&  2&\cdots&2& 
\end{array}\label{eq:Aquiver}
\eeq
where $\su(k)_L$ and $\su(k)_R$ are flavor symmetries and other $\su(k)$'s are gauge groups on the tensor branch.
We denote this conformal matter as $\CT_{k,N}^{6d}$.

\paragraph{Five dimensions.}
Following our general discussions of the previous section, we consider a 5d version of the quiver gauge theory of the form \eqref{eq:Aquiver}.
This is a 5d $\SU(k)^{N-1}$ quiver theory with $k$ flavors at each end, and the properties of this theory can be easily read off from the brane web construction
of this theory \cite{Aharony:1997ju,Aharony:1997bh,DeWolfe:1999hj} as a D5-NS5 system.
The theory has a UV fixed point which we denote as $\CS^{5d}_{k,N}$.
This 5d theory has global symmetry $\SU(k)_L \times \SU(k)_R \times \SU(N)_L \times \SU(N)_R$, where $\SU(N)_L \times \SU(N)_R$ is the enhanced symmetry.

The theory $\CS^{5d}_{k,N}$ itself is an SCFT, but by deforming it by mass term $m_{\SU(N)}$ 
in the Cartan of the diagonal subgroup of $\SU(N)_L \times \SU(N)_R$,
we get the IR $\SU(k)^{N-1}$ quiver theory 
\beq
\CS^{5d}_{k,N} \xrightarrow{\SU(N)~\text{mass deform}} [\SU(k)_L]-\SU(k)-\cdots-\SU(k)-[\SU(k)_R]. \label{eq:massdfSU}
\eeq
The gauge coupling is determined by the general formula \eqref{eq:quivercoup} which in this case is given by
$8\pi^2/g^2_i= m_{\SU(N),i}-m_{\SU(N),i+1}$ ($i=1,\cdots,N-1$), where $m_{\SU(N)}=\diag(\cdots ,m_{\SU(N),i}, \cdots)$.
This is precisely as expected from the brane construction of this theory.
Furthermore, this theory has a duality $ k \leftrightarrow N$ which can be readily seen from 
the brane construction.
Therefore, if we deform the theory by masses in the Cartan of the diagonal subgroup of $\SU(k)_L \times \SU(k)_R$,
we get the IR $\SU(N)^{k-1}$ quiver theory,
\beq
\CS^{5d}_{k,N} \xrightarrow{\SU(k)~\text{mass deform}} [\SU(N)_L]-\SU(N)-\cdots-\SU(N)-[\SU(N)_R], \label{eq:massdfsu}
\eeq
where $\SU(N)_{L, R}$ are flavor symmetries, and there are $k-1$ $\SU(N)$ gauge groups.

Now, our claim is that the compactification of the conformal matter $\CT_{k,N}^{6d}$ on $S^1$ is given by 
the theory $\CS^{5d}_{k,N}$ with the diagonal subgroup of $\SU(N)_L \times \SU(N)_R$ gauged,
\beq
\CT_{k,N}^{6d} \xrightarrow{S^1} \CS^{5d}_{k,N}\{ \SU(k)_L,\SU(k)_R, \SU(N)_L, \SU(N)_R \} / \SU(N)_{\rm diag} \label{eq:rd5dA}
\eeq
where the notation of the right hand side means that we are gauging the diagonal subgroup $\SU(N)_{\rm diag} \subset \SU(N)_L \times \SU(N)_R$
by the $\SU(N)$ vector multiplet.

Let us consider two types of deformation of this 5d theory. The first one is to go to the Coulomb branch of the $\SU(N)$ gauge group
by giving a vev to the adjoint scalar $\Phi_{\SU(N)}$. Then, this gives mass deformation of the theory $\CS^{5d}_{k,N}$,
and we exactly get the dimensional reduction of the 6d quiver \eqref{eq:Aquiver}.

Next, let us consider mass deformation of the diagonal subgroup of the flavor symmetry $\SU(k)_L \times \SU(k)_R$
at the origin of the Coulomb moduli space. This corresponds to introducing flavor Wilson lines on $S^1$.
In this case, the mass deformation of $\CS^{5d}_{k,N}$ is given by \eqref{eq:massdfsu}, but the diagonal subgroup of 
$\SU(N)_L \times \SU(N)_R$ is gauged by the gauge group $\SU(N)$ as in \eqref{eq:rd5dA}.
Therefore, we get an $\SU(N)^k$ necklace quiver theory.
This is exactly the one obtained by putting $N$ D4-branes on the $A_{k-1}$ singularity with generic $B$-flux.
In this way, two different 5d IR theories
follow from the single strongly interacting 5d SCFT $\CS^{5d}_{k,N}$.

\paragraph{Four dimensions.}
The $T^2$ compactification of the conformal matter $\CT_{k,N}^{6d}$ is now given as 
\beq
\CT_{k,N}^{6d} \xrightarrow{T^2} \CS^{4d}_{k,N}\{ \SU(k)_L,\SU(k)_R, \SU(N)_L, \SU(N)_R \} / \SU(N)^\tau_{\rm diag} \label{eq:rd4dA}
\eeq
where $\CS^{4d}_{k,N}$ is the 4d theory obtained by the $S^1$ compactification of $\CS^{5d}_{k,N}$,
and  the notation of the right hand side means that we are gauging the diagonal subgroup $\SU(N)_{\rm diag} \subset \SU(N)_L \times \SU(N)_R$
by the $\SU(N)$ vector multiplet with gauge coupling $\tau$.
Thus, the problem of $T^2$ compactification of the conformal matter $\CT_{k,N}^{6d}$ is reduced to the problem of 
$S^1$ compactification of $\CS^{5d}_{k,N}$.

Before going to study the 4d theory $\CS^{4d}_{k,N}$, we prepare some notation of class S theories \cite{Gaiotto:2009we,Gaiotto:2009hg}.
See \cite{Tachikawa:2015bga} for a review of class S theories and notations used in this paper.
We denote by $\sT_{k}\{Y_1, Y_2,Y_3\}$ the class S theory of type $A_{k-1}$ on a Riemann sphere with three punctures 
$Y_1$, $Y_2$ and $ Y_3$. These $Y$'s are specified by partition of $k$. For example,
a simple puncture is given by the partition of $k$ as $k=(k-1)+1$ and is denoted as $Y_{\rm simple}=[k-1, 1]$.
Similarly, a full puncture is given by the partition $k=1+\cdots+1$ and denoted as $Y_{\rm full}=[1^k]$.
More generally, the class S theory of type $A_{k-1}$ on a Riemann surface $C_{g,n}$ of genus $g$ with $n$ punctures $Y_1, \cdots, Y_n$
is denoted as $\sS_k\vev{C_{g,n}}\{Y_1,\cdots, Y_n\}$.

Now we study the theory $\CS^{4d}_{k,N}$. Because of the symmetry $k \leftrightarrow N$ of this theory, we assume $N \geq k$ for the moment. 
For the purpose of studying $\CS^{4d}_{k,N}$, we consider the mass deformation \eqref{eq:massdfSU} and \eqref{eq:massdfsu} in 4d.
The right hand side of \eqref{eq:massdfSU} is a class S theory of $A_{k-1}$ type on a Riemann sphere
with two full punctures and $N$ simple punctures.
As discussed above, the gauge couplings are determined by the mass deformation.
Then, by tuning the mass deformation, we can collide the $N$ simple punctures at a single point and obtain \cite{Gaiotto:2009we},
\beq
\sT_{k}\{ [1^k], [1^k], [1^k] \} - \SU(k)-\cdots-\SU(k)-\SU(k-1)-\cdots-\SU(1),~~(N\geq k) \label{eq:partdefm1}
\eeq
where there are $N-k+1$ $\SU(k)$'s, and
each gauge group is coupled to additional fundamentals if necessary so that the gauge group becomes conformal.
The $\SU(1)$ is introduced formally.
The leftmost $\SU(k)$ is coupled to one of the full punctures of $\sT_{k}\{ [1^k], [1^k], [1^k] \}$.
On the other hand, the right hand side of \eqref{eq:massdfsu} is a class S theory of type $A_{N-1}$ on a Riemann sphere
with two full punctures and $k$ simple punctures. Then, by tuning the masses and colliding simple punctures, we get (when $N \geq k$),
\beq
\sT_{N}\{ [1^N], [1^N], [N-k,1^k] \} - \SU(k)-\SU(k-1)-\cdots-\SU(1),~~(N \geq k) \label{eq:partdefm2}
\eeq 
where $\SU(k)$ is coupled to the puncture $[N-k,1^k]$.

From the above results, we expect that the theory $\CS^{4d}_{k,N}$ contains both of the theories
$\sT_{k}\{ [1^k], [1^k], [1^k] \}$ and $\sT_{N}\{ [1^N], [1^N], [N-k,1^k] \} $ when $N \geq k$. We propose that this theory is given by
\beq
\CS^{4d}_{k,N} =\left\{ \begin{array}{ll}
\sT_{N}\{ [1^N], [1^N], [N-k,1^k] \}  -\SU(k)-\sT_{k}\{ [1^k], [1^k], [1^k] \}  & (N >k) \\[0.3cm]
\sT_{N}\{ [1^N], [1^N], [1^N] \}  -[\SU(N)+\text{one fund.}]-\sT_{N}\{ [1^N], [1^N], [1^N] \}  & (N=k) \\[0.3cm]
\sT_{N}\{ [1^N], [1^N], [1^N] \}- \SU(N)- \sT_{k}\{ [1^k], [1^k], [k-N,1^N] \} & (N<k)
\end{array} \right. \label{eq:4dS}
\eeq
where in the $N=k$ case there is one fundamental representation coupled to the middle $\SU(N)$ gauge group.\footnote{In the notation of the introduction, for $\CT^\text{6d}=\CT^\text{6d}_{k,N}$,
$\CS^\text{4d}=\CS^\text{4d}_{k,N}$,
and $\cU=\sT_N\{\cdots\}$, $\cV=\sT_k\{\cdots\}$. When $N=k$, one additional fundamental needs to be included in either $\cU$ or $\cV$.}

The contribution of the $\sT_{N}\{ [1^N], [1^N], [N-k,1^k] \}$ theory to the beta function of the $\SU(k)$ is the same as
that of $k+1$ fundamentals when $k<N$. 
So in each case, the gauge group $\SU(k)$ or $\SU(N)$ appearing in the above equation has IR free beta function.
We will give other justifications of the appearance of the IR free gauge group later in this paper.

We will give more checks of \eqref{eq:4dS} below, but before doing that, 
let us complete our task of determining the 4d theory obtained by compactification of the 6d conformal matter $\CT_{k,N}^{6d}$.
The 4d theory is obtained by gauging the diagonal subgroup $\SU(N)_{\rm diag} \subset \SU(N)_L \times \SU(N)_R$ of the $\CS^{4d}_{k,N} $.
This can be easily done in the class S theory. We just need to replace $\sT_{N}\{ [1^N], [1^N], Y \}  $ ($Y=[N-k,1^k] $ or $[1^N]$)
by the theory on a torus $\sS_N \vev{T^2_\tau}\{ Y \}$,  
where we used the class S notation introduced above. The $T^2_\tau$ is a torus with the complex modulus $\tau$.
Therefore, the final result is
\beq
\CT_{k,N}^{6d}  \xrightarrow{ T^2 } 
\left\{ \begin{array}{ll}
\sS_N \vev{T^2_\tau}\{ [N-k,1^k] \} -\SU(k)-\sT_{k}\{ [1^k], [1^k], [1^k] \}  & (N >k) \\[0.2cm]
\sS_N \vev{T^2_\tau}\{ [1^N]  \}  -[\SU(N)+\text{one fund.}]-\sT_{N}\{ [1^N], [1^N], [1^N] \}  & (N=k) \\[0.2cm]
\sS_N \vev{T^2_\tau}\{ [1^N]  \} - \SU(N)- \sT_{k}\{ [1^k], [1^k], [k-N,1^N] \} & (N<k)
\end{array} \right. \label{eq:4dconfm}
\eeq
This theory has the $\SL(2,\bZ)$ S-duality group acting on $\sS_N \vev{T^2_\tau}\{ [1^N]  \}$, and 
has manifest $\SU(k)_L \times \SU(k)_R$ flavor symmetry from the two full punctures $[1^k]$. 

To give further checks of the above proposal, we need a mass deformation of the theory
$\sT_N\{[1^N], [1^N],Y\}$.
The following facts are known \cite{Bergman:2014kza,Hayashi:2014hfa}. 
(See also \cite{Benini:2009gi,Bao:2013pwa,Hayashi:2013qwa,Aganagic:2014oia,Hayashi:2014wfa,Hayashi:2015xla}.)
The following statements hold in both 4d and 5d versions of the theory $\sT_N\{[1^N], [1^N],Y\}$.

Let us give generic masses to the diagonal subgroup of $\SU(N)_L \times \SU(N)_R $ of the full punctures.
Then this theory flows in the IR to a linear quiver
\beq
\sT_N\{[1^N], [1^N],Y\}     \xrightarrow{\SU(N)_{\rm diag}~ \text{mass deform}} \SU(v_1)-\SU(v_2)-\cdots-\SU(v_{N-1}) \label{eq:massRGflow}
\eeq
In this quiver, each gauge group is coupled to additional fundamentals if necessary so that each gauge group becomes conformal.
The $v_i $ are determined as follows. The $Y$ is specified by a partition of $N$ as $Y=[m_1,m_2,\cdots, m_\ell]$.
This partition $Y$ defines a Young diagram. Then we can consider the transpose of the Young diagram $Y$,
which we denote as $Y^T=[n_1, \cdots, n_k]$ where $n_1 \geq \cdots \geq n_k$. We also define $n_i=0$ for $i>k$.
Then $v_i$ is determined by
\beq
v_{i-1} - v_{i} =1-n_i ,~~~v_{N-1}=1. \label{eq:generalmdtn}
\eeq

If $Y$ is given by $Y=[N-k,1^k]$ with $N>k$, then $Y^T=[k+1, 1^{N-k-1}]$ and hence $n_1=k+1$, $n_i=1$ for $2 \leq i \leq N-k$ and $n_i=0$ for $i >N-k$.
Then $v_i=k$ for $i \leq N-k$ and $v_i=N-i$ for $N-k \leq i  \leq N-1 $, and the quiver becomes
\beq
[\SU(k)]-\SU(k)-\cdots - \SU(k)-\SU(k)-\cdots-\SU(1).\label{eq:mdtn1}
\eeq
The $[\SU(k)]$ is a flavor symmetry coming from the fundamentals coupled to the leftmost $\SU(k)$. This $[\SU(k)]$ is identified with the flavor symmetry
of the puncture $Y=[N-k,1^k]$.
There are $N-k$ $\SU(k)$ gauge groups.
Similarly, if $Y=[1^N]$ we get
\beq
[\SU(N)]-\SU(N-1)- \SU(N-2) - \cdots-\SU(1).\label{eq:mdtn2}
\eeq

Now we can discuss mass deformation of $\CS^{4d}_{k,N}$ in \eqref{eq:4dS}.
Let us mass-deform the diagonal subgroup of $\SU(N)_L \times \SU(N)_R$ in \eqref{eq:4dS}.
When $N \geq k$, by using \eqref{eq:mdtn1} one can see that we precisely get the theory \eqref{eq:partdefm1}.
Similarly, if we deform the $\SU(k)_L \times \SU(k)_R$ in \eqref{eq:4dS}, then by using \eqref{eq:mdtn2} with $N$ replaced by $k$, 
we precisely get the theory \eqref{eq:partdefm2}. 
This gives a strong check of our proposal \eqref{eq:4dS}.
In particular, note that the IR free gauge group appearing in \eqref{eq:4dS} becomes
conformal after the mass deformation of either $\SU(N)$ or $\SU(k)$.
The conformality of gauge groups after the deformation of $\SU(N)$ was indeed shown in our general discussion of the previous section from the 6d point of view.

We have seen that \eqref{eq:partdefm1} and \eqref{eq:partdefm2} can be obtained by mass deformation of $\SU(N)$ and $\SU(k)$ in \eqref{eq:4dS},
respectively. By going back the duality, we can also get the 4d version of the right hand side of \eqref{eq:massdfSU} and \eqref{eq:massdfsu},
respectively. In the compactification of $\CT_{k,N}^{6d} $, the diagonal subgroup of $\SU(N)_L \times \SU(N)_R$ is gauged.
In this way, we get two theories; one is a linear $\SU(k)^{N-1}$ quiver with the gauge coupling determined by the vev of $\Phi_{\SU(N)}$,
and the other is a necklace $\SU(N)^{k}$ quiver. These are the theories discussed in \cite{DelZotto:2015rca}.
Now we can see that these two theories flow from the single 4d theory \eqref{eq:4dconfm} which has manifest $\SL(2,\bZ)$ S-duality and
$\SU(k)_L \times \SU(k)_R$ flavor symmetry.

\subsection{M-theory interpretation}\label{app:4dconfm}
Here we try to understand \eqref{eq:4dconfm} in terms of M5 branes in M-theory.
As mentioned above, the A-type conformal matter is realized in M-theory by putting $N$ coincident M5 branes on $A_{k-1}$ singularity.
If we realize this $A_{k-1}$ singularity by Taub-NUT space and go to type IIA string theory, we get a system
of $N$ coincident NS5 branes and $k$ coincident D6 branes intersecting with each other.
The A-type conformal matter is realized on the intersection.

Now we compactify the theory on $T^2$ so that we get a $T^2$ compactification of the conformal matter.
Taking T-dual twice, we get $N$ coincident NS5 branes and k coincident D4 branes.
Uplifting to M-theory, we get $N$ coincident M5 branes and $k$ coincident M5 branes intersecting on dimension 4 subspace.

The directions in which M5 branes are extending after the above duality chain are listed in table~\ref{tab:Mdirection}.
They are intersecting on the space $\bR^{1,3}$ .
Furthermore, $N$ M5 branes are compactified on $T^2$, and $k$ M5 branes are compactified on $S^1 \times \bR$.

Let us focus on the $N$ M5 branes. This is compactified on $T^2$, so it is a class S theory of $A_{N-1}$ type on $T^2$.
From the point of view of this $N$ M5 branes, the $k$ M5 branes look like a codimension 2 defect, and hence it is a kind of puncture.
So it is natural to obtain a theory $\sS_N\vev{T^2_\tau}\{ Y \}$, where $Y$ is specified by the $k$ M5 branes.
Next, let us focus on the $k$ M5 branes. This is compactified on $S^1 \times \bR$, but this space can be regarded as a sphere
with two full punctures in class S theory. So this is a class S theory of type $A_{k-1}$ on a Riemann sphere with two full punctures
and one puncture $Y'$ specified by the $N$ M5 branes which look like a puncture from the point of view of the $k$ M5 branes.
Thus we get the theory $\sT_k\{[1^k],[1^k], Y'\}$.
These observations partly explain the structure of \eqref{eq:4dconfm}.
Conversely, our results tell us what exactly happens in this setup of M5 branes.

When $N=1$, one M5 brane is a simple puncture from the point of view of the $k$ M5 branes \cite{Gaiotto:2009gz}.
This was also found in minimal conformal matters of general ADE type \cite{Ohmori:2015pua}.
Our result is consistent with this because in this case $[k-N,1^N]=[k-1,1]$ is a simple puncture.

\begin{table}[t]
\centering
\begin{tabular}{|c|c|c|c|c|}
\hline
& $\bR^{1,3}$ &  $T^2$ (or $S^1 \times \bR$) & $S^1 \times \bR$ & $\bR^3$ \\ \hline
$N$ M5 branes & $ \bullet$ & $\bullet$  & & \\ \hline
$k$ M5 branes & $ \bullet  $ & & $\bullet$ & \\ \hline
\end{tabular}
\caption{Directions in which M5 branes extend.}\label{tab:Mdirection}
\end{table}

It is also clear that if we replace the $T^2$ of table~\ref{tab:Mdirection} by $S^1 \times \bR$,
the theory we obtain from the M5 branes' intersection should be $\CS^{4d}_{k,N} $ in \eqref{eq:4dS}.
This is a little progress in the understanding of M-theory and \Nequals{(2,0)} theory. In general, it is very interesting to study 
what happens when two bunches of M5 branes intersect with each other along dimension 4 subspace.
This is a difficult problem to answer if the M5 branes are intersecting in flat $\bR^{1,10}$ space, because the \Nequals{(2,0)} theory
is intrinsically strongly coupled and hence there is no clear separation between the bulk \Nequals{(2,0)} theory and 
the 4d theory living on the intersection. 
However, if we compactify the M5 branes on $S^1$, we get 5d \Nequals2  super Yang-Mills
which is weakly coupled in the IR limit. Then it becomes a well-defined question to ask what theory is living on the intersection.
If we compactify the system on $S^1$ which is common to both $N$ M5 branes and $k$ M5 branes, the system is reduced
to a well-known situation in which D4 branes are intersecting and we just get free hypermultiplets in 3d.
Instead, if we compactify the system on two $S^1$'s as in table~\ref{tab:Mdirection} with the replacement $T^2 \to S^1 \times \bR$, 
the intersection looks like a codimension one domain wall
from the point of view of each of the 5d \Nequals2  super Yang-Mills theories.
What we found is that the theory living on this domain wall is
the 4d theory $\CS^{4d}_{k,N} $ in \eqref{eq:4dS}.
Flavor symmetries $\SU(N)_L \times \SU(N)_R$ and $\SU(k)_L \times \SU(k)_R$ are naturally coupled to
the gauge groups of 5d $\SU(N)$ and $\SU(k)$ \Nequals2  super Yang-Mills theories on the two sides of the domain walls, respectively.

\subsection{Generalization}\label{sec:GT}
We can easily generalize the results of Sec.~\ref{sec:Aconf} to more general theories of \cite{Gaiotto:2014lca}. 
These theories are obtained from the theory \eqref{eq:Aquiver} in the following way. In this subsection we assume $N \geq k$.

We take two partitions of $k$, denoted as $Y_{L}=[m_i^{L}]$ and $Y_{R}=[m_i^{R}]$ ($m _i^{L,R} \geq m_{i+1}^{L,R}$).
They define homomorphisms $\rho_{L,R}:\SU(2) \to \SU(k)$ such that the $k$ dimensional representation of $\SU(k)$
is decomposed into $m_i^{L,R}$ dimensional representations of $\SU(2)$. (Note $k=\sum_i m_i^{L,R}$ by definition.)
Then we can define nilpotent elements of $\SU(k)$ as $\rho_{L,R}(\sigma^+)$, where $\sigma^+$ is the raising operator of $\SU(2)$.
Let $\mu_L$ and $\mu_R$ be the holomorphic moment maps (or simply meson chiral operators) associated to the flavor symmetries
$\SU(k)_L$ and $\SU(k)_R$ respectively. Using the nilpotent elements, we can higgs the theory \eqref{eq:Aquiver} by giving vevs to the
moment maps as $\mu_{L,R} \propto \rho_{L,R}(\sigma^+)$. Then in the low energy, we get a new theory with some decoupled free fields
such as Nambu-Goldstone bosons associated to the symmetry breaking of $\SU(k)_{L,R}$.
We denote the interacting part of the theory as $\CT^{6d}_{k,N}\{Y_L,Y_R \}$.
When $Y_L=Y_R=[1^k]$, this is just the original theory \eqref{eq:Aquiver}.

The result of the higgsing is (see \cite{Tachikawa:2013kta} and references therein),
\beq
\begin{array}{ccccccccc}
 & \su(u_1) & \cdots & \su(u_{N-1}) &  \\
&  2&\cdots&2& 
\end{array}\label{eq:GTquiver}
\eeq
where each gauge group is coupled to additional fundamentals if necessary to make them conformal, 
and $u_i~(i=1,\cdots, N-1)$ are determined as follows. Let $Y_{L,R}^T=[n_i^{L,R}]$ be the partitions of $k$ obtained by
taking the transpose of the Young diagrams associated to $Y_{L,R}$. We formally take $n_i^{L,R}=0$ for large $i$ which does not appear in $Y_{L,R}$.
Then $u_i$ is given as
\beq
u_{i-1} - u_{i} = n^L_{N+1-i} - n^R_{i},~~~u_{0}, u_{N} :=0.
\eeq
In this way, the generalized theories $\CT^{6d}_{k,N}\{Y_L,Y_R \}$ are defined.

As already discussed in the general arguments of the previous section, the 5d version of the quiver \eqref{eq:GTquiver}
is expected to have a UV fixed point $\CS^{5d}_{k,N} \{Y_L,Y_R\}$ with enhanced $\SU(N)_L \times \SU(N)_R$ symmetry.
Then the $S^1$ compactification of $\CT^{6d}_{k,N}\{Y_L,Y_R \}$ is given by this $\CS^{5d}_{k,N} \{Y_L,Y_R\}$ with the diagonal
subgroup of $\SU(N)_L \times \SU(N)_R$ gauged.

It is also easy to determine the 4d theory. We just need to higgs the moment maps $\mu_L$ and $\mu_R$ of the theory \eqref{eq:4dconfm}
by nilpotent vevs. The result is
\beq
\CT_{k,N}^{6d}\{Y_L, Y_R\}  \xrightarrow{ T^2 } 
\left\{ \begin{array}{ll}
\sS_N \vev{T^2_\tau}\{ [N-k,1^k] \} -\SU(k)-\sT_{k}\{ [1^k], Y_L, Y_R \}  & (N >k) \\[0.2cm]
\sS_N \vev{T^2_\tau}\{ [1^N]  \}  -[\SU(N)+\text{one fund.}]-T_{N}\{ [1^N], Y_L, Y_R \}  & (N=k) \\[0.2cm]
\sS_N \vev{T^2_\tau}\{ [1^N]  \} - \SU(N)- \sT_{k}\{ [k-N,1^N],Y_L, Y_R \} & (N<k)
\end{array} \right. \label{eq:4dGT}
\eeq
This is the $T^2$ compactification of the class of theories in \cite{Gaiotto:2014lca}.

\subsection{Cases without IR-free gauge group}\label{sec:withoutIR-A}
There is actually a special subclass of theories in which the IR free gauge group does not appear.
We take $k=N$ and $Y_L=[N]$ ($Y_L^T=[1^N]$). For simplicity, let us first consider the case $Y_R=[1^N]$ ($Y_R^T=[N]$).
Then the 6d theory is given by
\beq
\begin{array}{ccccccccc}
 & \su(N-1) & \cdots & \su(2) & \su(1)  \\
&  2&\cdots&2& 2
\end{array}
+ \text{one fund.\ of flavor $\su(N)$},
\eeq
where additional free hypermultiplet can be seen from the type IIA construction.
Such a non-interacting hypermultiplet charged under the remaining flavor symmetry exists for any $Y_R$, and we call the interacting part $\CT_{N,N}^{\text{6d}}\{[N],Y_R\}\intpart$.
In the 4d theory, one of the punctures $Y_L$ is completely higgsed and this puncture disappears. It is called the closing of the puncture.
After this, we get a theory $\sT_N[[1^N], [1^N]]$ with two full punctures, or equivalently a theory on a tube (with Dirichlet boundary conditions
at the two ends when the \Nequals{(2,0)} theory is reduced to 5d \Nequals2  super Yang-Mills). 
This theory is actually not an interacting SCFT.
The $\SU(N) \times \SU(N)_R$ symmetries associated to the full punctures are automatically broken down to the diagonal subgroup \cite{Gaiotto:2011xs}.
Then, when the $\SU(N)$ is gauged, the gauge group is completely higgsed by this theory $\sT_N[[1^N], [1^N]]$ and only the flavor $\SU(N)_R$
survives by mixing with the gauge group. By applying these facts to \eqref{eq:4dGT}, we get
\beq
\CT_{N,N}^{6d}\{[N], [1^N]\} \xrightarrow{ T^2 } \sS_N \vev{T^2_\tau}\{ [1^N]  \}+\text{one fund.}
\eeq
Here, one can check that there are $N$ free decoupled hypermultiplets in 4d after the process of nilpotent higgsing as can be checked
by the method of \cite{Tachikawa:2013kta}, and these decoupled hypermultiplets are identified with the additional hypermultiplets in \eqref{eq:4dGT}
in the fundamental representation of $\SU(N)$ which is higgsed.
Subtracting the hypermultiplets form both side, we get
\beq
\begin{array}{ccccccccc}
 & \su(N-1) & \cdots & \su(2) & \su(1)  \\
&  2&\cdots&2& 2
\end{array}
\xrightarrow[\text{point}]{\text{conformal}}
\CT_{N,N}^{6d}\{[N], [1^N]\}\intpart \xrightarrow{ T^2 } \sS_N \vev{T^2_\tau}\{ [1^N]  \}.
\eeq

In the same way, we can also consider general $Y_R:=Y$. The interacting part of the 6d theory is
\beq
\begin{array}{ccccccccc}
 & \su(v_1) & \cdots & \su(v_{N-1}) &  \\
&  2&\cdots&2& 
\end{array}\label{eq:hty}
\eeq
where $v_i$ are defined by \eqref{eq:generalmdtn}. 
Note that $v_{N-1}=1$.
We can simply partially close $[1^N]$ in the above equation to obtain
\beq
\begin{array}{ccccccccc}
 & \su(v_1) & \cdots & \su(v_{N-1}) &  \\
&  2&\cdots&2& 
\end{array}
\xrightarrow[\text{point}]{\text{conformal}}
\CT_{N,N}^{6d}\{[N], Y  \}\intpart  \xrightarrow{ T^2 } \sS_N \vev{T^2_\tau}\{ Y  \} \label{eq:4dhty}
\eeq
for arbitrary $Y$. In this class of theories, the corresponding 4d theory is conformal without any IR free gauge group.

We can also derive the above results much more directly. As already described in Sec.~\ref{sec:Aconf},
the 5d version of the quiver \eqref{eq:hty} has a fixed point which is a 5d version of the $\sT_N$-like theory,
$\sT^{5d}_N\{ [1^N], [1^N],Y \}$. Thus in our notation above, we find that $\CS^{5d}_{N,N} \{[N],Y\} =\sT^{5d}_N\{ [1^N], [1^N],Y \} $.
The $S^1$ compactification of $\CT_{N,N}^{6d}\{[N], Y  \}\intpart$ is thus the $\sT^{5d}_N\{ [1^N], [1^N],Y \} $ theory with the diagonal
subgroup of $\SU(N)_L \times \SU(N)_R$ coming from the full punctures gauged.
By reducing this theory further to 4d, we immediately get \eqref{eq:4dhty}.

\section{Conformal matters and class S theories, general type}\label{sec:examples2}
Next, let us discuss the 6d theory $\CT_{N}\{\fg,\fg\}$ on the worldvolume of $N$ M5-branes on $\bC^2/\Gamma_\fg$ singularity, where $\fg$ can be  $D_k$ or $E_k$.
The authors have not been able to obtain as full an answer as in the case of $\fg=A_{k-1}$,
but we can still understand quite a lot. 
Also, even for $\fg=A_{k-1}$, the analysis in this section sheds some new light.

\subsection{Structure of the 5d reduction}
On the tensor branch in 6d, the quiver is of the form 
\beq
\begin{array}{ccccccccc}
[\g] & \g & \cdots & \g &[ \g] \\
&  2&\cdots&2& 
\end{array}
\eeq
where the bifundamental `matter' of $\fg\times \fg$ is a nontrivial 6d very Higgsable SCFT.

First let us compactify on $S^1$ without any Wilson line.
From our general discussion, its $S^1$ compactification is given by a 5d $\SU(N)$ gauge theory coupled to a strongly-coupled SCFT $\CS^{5d}\{\fg,\fg,\SU(N)\}$,
which is the strongly-coupled SCFT limit of the 5d quiver
\beq
[\g_L] - \g - \cdots - \g - [\g_R],\label{eq:qv1}.
\eeq
where bifundamentals are nontrivial 5d conformal theories.
To the knowledge of the authors, no study has been done on such quivers with generalized matters in 5d, but our general discussion in Sec~\ref{sec:general} requires that
there is an enhancement of the flavor symmetry of \eqref{eq:qv1}
from $\U(1)^{N-1}$ instanton symmetries to $\SU(N)$, 
just as in the case when $\g$ is of type $A$ where the matter fields are free bifundamental hypermultiplets.

The same 5d SCFT $\CS^{5d}\{\fg,\fg,\SU(N)\}$ can be identified as follows.
If we instead compactify the 6d theory on $S^1$ with generic Wilson lines in the diagonal subgroup of the flavor symmetry $\g_L \times \g_R$,
we get a 5d ordinary quiver theory whose nodes form the affine Dynkin diagram of type $\g$. The gauge group is \begin{equation}
\prod_{a=0}^{\rank \fg} \SU(d_a N)
\end{equation} where $d_0=1$ corresponds to the affine node
and the vector $(d_a)$ is in the kernel of the affine Cartan matrix. 
There is as always the bifundamental matter fields for the edges of the Dynkin diagram.

This is the theory realized by putting $N$ D4 branes on singularities of type $\g$ in type IIA string theory with generic $B$-flux through
the singularities. 
In this description, the gauge group $G=\SU(N)$ is manifestly visible: it is located at the affine node $a=0$ of the affine Dynkin diagram. 
If we remove this $G$ vector multiplet, the rest of the quiver is of the form
of a finite Dynkin diagram which is expected to have a UV fixed point. In fact, in type IIA, the (weighted) sum of 
inverse gauge coupling squared is constrained by string coupling and $\alpha'$, but other linear combinations can be changed 
by changing the $B$-flux. By turning off the $B$-flux, the gauge couplings of the gauge groups located at the finite Dynkin diagram
can be taken to be infinity, implying the existence of the 5d fixed point theory $\CS^{5d}\{\fg,\fg,\SU(N)\}$.
The remaining $\SU(N)$ at the extended node is our $G$ vector multiplet of the general discussion.

In summary, we have two theories. One is the theory \eqref{eq:qv1} and the other is the theory
\beq
\text{finite Dynkin quiver of type $\g$ with the gauge group $\prod_{a=1}^{\rank \fg}\SU(d_a N)$}. \label{eq:qv2}
\eeq
These theories \eqref{eq:qv1} and \eqref{eq:qv2} should have a common UV fixed point $\CS^{5d}\{\fg,\fg,\SU(N)\}$, with the flavor symmetry $\g_L \times \g_R \times \SU(N)$.
Only $\g_L\times \g_R$ is manifest in \eqref{eq:qv1},
which is obtained by mass deformation in $\SU(N)$ of $\CS^{5d}\{\fg,\fg,\SU(N)\}$,
while only $\SU(N)$ is manifest in \eqref{eq:qv2} which is obtained by mass deformation in the diagonal subgroup of $\g_L \times \g_R$.
In this sense, these two IR theories \eqref{eq:qv1} and \eqref{eq:qv2} are dual to each other.
This is the precise version of the ``novel 5d duality'' of \cite{DelZotto:2014hpa}.
The case of $N=1$ and $\g=D_n$ was studied explicitly in \cite{Ohmori:2015pua}.

Summarizing, the compactification on $S^1$ of the 6d theory$\CT^\text{6d}_N\{\fg,\fg\}$ has the structure shown in Fig.~\ref{fig:big1}.  The 5d theory becomes a generalized quiver on the part of the 5d Coulomb branch that corresponds to the 6d tensor branch,
and becomes a standard affine quiver when mass deformed. 

\begin{figure} 
\[
\begin{tikzpicture}
\node (6d1) at (0,3) {$\CT^\text{6d}_N\{\fg,\fg\}$};
\node (5d1) at (0,1) {$\CS^\text{5d}\{\fg,\fg,\SU(N)\}/ \SU(N)_{R^{-1}}$};
\node[align=center] (4d1a) at (-4,-1) {generalized linear quiver \\ with gauge group $\fg^{N-1}$\\ with $\fg\times\fg$ flavor symmetry};
\node[align=center] (4d1b) at (4,-1) {standard affine quiver \\ with gauge group $\prod_{a=0}^{\rank \fg} \SU(d_aN)$} ;
\draw[->] (6d1)--node[anchor=east]{$S^1_R $} (5d1);
\draw[->,in=90] (5d1)-- node[anchor=east]{C. branch} (4d1a.north);
\draw[->,in=-90] (5d1)--node[anchor=west]{mass deform.} (4d1b.north);
\end{tikzpicture}.
\]
\caption{$S^1$ reductions of the theory  $\CT^\text{6d}_N\{\fg,\fg\}$.\label{fig:big1}}
\end{figure}

\subsection{An aside: T-duality of \Nequals{(1,0)} little strings}\label{sec:little}
Here let us pause for a moment and describe another way to embed this ``novel 5d duality'' of $\CS^{5d}\{\fg,\fg,\SU(N)\}$ into a 6d theory.  
For this purpose, note that the 6d version of the quiver theory \eqref{eq:qv2} can be realized in F-theory by having $\rank\fg$ ($-2$)-curves intersecting according to the Dynkin diagram of type $\fg$,
and by decorating them by 7-branes to give the gauge groups $\SU(d_aN)$.  There is one infinitely-large flavor curve supporting $\SU(N)$ flavor symmetry which corresponds to the affine node. When $\fg$ is of type $A$, this flavor curve intersects with two compact curves, thus the flavor symmetry is in fact $\SU(N)\times \SU(N)$, but in the following we mostly consider only the diagonal subgroup $\SU(N)_\text{diag}$.

Let us denote the 6d SCFT at the origin of this tensor branch as $\CT^\text{6d}_\fg\{\SU(N)\}$.  From our general discussion in Sec.~\ref{sec:general},
its $S^1$ compactification is given by the theory $\CS^\text{5d}\{\fg,\fg,\SU(N)\}/\fg_\text{diag}$.
When we deform this setup by the $\SU(N)$ mass terms, 
what we find is the 5d quiver  \eqref{eq:qv1} whose diagonal $\fg$ flavor symmetry is gauged by 5d $\fg$ vector multiplet. 
Stated differently, this is a $\fg^N$ circular quiver with generalized bifundamental ``matter'' contents. 
Summarizing, we have the situation given in Fig.~\ref{fig:big2}. 

\begin{figure}
\[
\begin{tikzpicture}
\node (6d1) at (0,3) {$\CT^\text{6d}_\fg\{\SU(N)\}$};
\node (5d1) at (0,1) {$\CS^\text{5d}\{\fg,\fg,\SU(N)\}/\fg_{\text{diag}, R^{-1}}$};
\node[align=center] (4d1a) at (-4,-1)  {generalized circular quiver \\ with gauge group $\fg^N$};
\node[align=center] (4d1b) at (4,-1){standard finite quiver \\ with gauge group $\prod_{a=1}^{\rank \fg} \SU(d_aN)$\\ with $\SU(N)$ flavor symmetry} ;
\draw[->] (6d1)--node[anchor=east]{$S^1_{R} $} (5d1);
\draw[->,in=90] (5d1)-- node[anchor=east]{mass deform.} (4d1a.north);
\draw[->,in=-90] (5d1)--node[anchor=west]{C. branch} (4d1b.north);
\end{tikzpicture}.
\]
\caption{$S^1$ reductions of the theory $\CT^\text{6d}_\fg\{\SU(N)\}$.\label{fig:big2}}
\end{figure} 

The two diagrams shown in Fig.~\ref{fig:big1} and Fig.~\ref{fig:big2} are almost symmetric, which can be understood using the T-duality of two 6d \Nequals{(1,0)} theories. 
On one side, consider the little string theory $\CT^\text{A}_{\fg,N}$ obtained from type IIA theory with $N$ NS5-branes on $\bC^2/\Gamma_\fg$.
Lifting to M-theory, this can be considered as the $\CT^\text{6d}_N\{\fg,\fg\}$ theory coupled to 6d $\fg$ gauge field supported on the $\bC^2/\Gamma_\fg$ singularity around the M-theory circle. 

On the other side, consider the little string theory $\CT^\text{B}_{\fg,N}$ obtained from type IIB theory with $N$ NS5-branes on $\bC^2/\Gamma_\fg$. 
By taking the S-duality, this is given by $N$ D5-branes on $\bC^2/\Gamma_\fg$, and therefore on a generic point on its tensor branch, this is given by the quiver $\prod^{\rank \fg}_{a=0} \SU(d_aN)$.  Equivalently, this little string theory is given by $\CT^\text{6d}_\fg\{\SU(N)\}$ coupled to an $\SU(N)$ vector multiplet. 

These two theories $\CT^\text{A}_{\fg,N}$ and $\CT^\text{B}_{\fg,N}$ are clearly T-dual.
Let $\alpha'_\text{6d}$ be the intrinsic length scale squared of the 6d theories $\CT^\text{A}_{\fg,N}$ and $\CT^\text{B}_{\fg,N}$.
In the string theory realization, it is the same as the usual $\alpha'$, but we attached the subscript ``6d" to emphasize that our statement 
is valid in purely six dimensional theories to the extent that the little strings are UV complete. 
The 6d gauge coupling $8\pi^2/g^2$ in each of the theories, which is either $\g_\text{diag}$ or $\SU(N)$, is given by $1/(2\pi \alpha'_\text{6d})$.
If we compactify $\CT^\text{A}_{\fg,N}$ on a circle $S^1$ with radius $R$, the gauge coupling of $\SU(N)$ is given by $R^{-1}$
while the gauge coupling of $\g_\text{diag}$ is given by $\tilde{R}^{-1}$, where we have defined $\tilde{R}:=\alpha'_\text{6d}/R$.
Its T-dual setup is the little string $\CT^\text{B}_{\fg,N}$ compactified on a different circle $\tilde S^1$ with radius $\tilde R$. 
Then we obtain the same 5d theory, which is the theory $\CS^\text{5d}\{\fg,\fg,\SU(N)\}$ coupled to the $\fg_\text{diag}$ gauge multiplet with gauge coupling $8\pi^2/g^2=\tilde R^{-1}$, and to the $\SU(N)$ gauge multiplet with gauge coupling $8\pi^2/g^2= R^{-1}$.

Combined, we see that the following picture arises, shown in Fig.~\ref{fig:Tdual}.
This is the system that explains the extreme similarity of the Figures~\ref{fig:big1} and \ref{fig:big2}.
\begin{figure}
\[
\begin{tikzpicture}
\node (6d1) at (-2,3) {$\CT^\text{A}_{\fg,N}$};
\node (6d2) at (2,3) {$\CT^\text{B}_{\fg,N}$};
\node (5d1) at (0,1) {$\CS^\text{5d}\{\fg,\fg,\SU(N)\}/\fg_{\text{diag},\tilde R^{-1}}\times \SU(N)_{R^{-1}}$};
\node[align=center] (4d1a) at (-4,-1) {generalized circular quiver \\ with gauge group $\fg^N$};
\node[align=center] (4d1b) at (4,-1) {standard affine quiver \\ with gauge group $\prod_{a=0}^{\rank \fg} \SU(d_aN)$} ;
\draw[->] (6d1)--node[anchor=east]{$S^1_R $} (5d1);
\draw[->] (6d2)--node[anchor=west]{$\tilde S^1_{\tilde R}$}(5d1);
\draw[<->,dotted] (6d1)--node[above]{T dual} (6d2);
\draw[->,in=90] (5d1)-- node[anchor=east]{C. branch} (4d1a.north);
\draw[->,in=-90] (5d1)--node[anchor=west]{C. branch} (4d1b.north);
\end{tikzpicture}.
\]
\caption{$S^1$ reductions of two T-dual theories $\CT^\text{A}_{\fg,N}$ and $\CT^\text{B}_{\fg,N}$.\label{fig:Tdual}}
\end{figure}

\subsection{Structure of the 4d reduction}\label{sec:notdetailedS}
Now let us compactify one further dimension and identify $\CS^\text{4d}\{\fg,\fg,\SU(N)\}$.
The question can be approached either from the point of view of the theory \eqref{eq:qv1} or \eqref{eq:qv2}. 
Here we choose to use \eqref{eq:qv1}. 

The deformation of $\CS^\text{4d}\{\fg,\fg,\SU(N)\}$ by the mass parameter for $\SU(N)$ is  
the 4d quiver  \begin{equation}
[\g_L] - \g - \cdots - \g - [\g_R] \label{eq:qv1-4d}.
\end{equation} where the generalized bifundamental $\sB_\fg$ of $\fg\times \fg$ 
comes from the $T^2$ reduction of the very Higgsable SCFT in 6d.
As studied in \cite{Ohmori:2015pua}, this generalized bifundamental
is given by a class S theory $\sB_\fg:=\sT_\fg\{\g,Y_\text{simple},\g \}$, i.e.~the class S theory  of type $\fg$ on a sphere with two full punctures  and a simple puncture.
Therefore, the quiver \eqref{eq:qv1-4d} theory itself is a class S theory of type $\fg$
on a sphere with two full punctures and $N$ simple punctures, which we denote as \begin{equation}
\sT_\fg\{\g,Y_\text{simple},\ldots,Y_\text{simple},\g\}
\end{equation} 
The $N-1$  cross ratios are the IR remnant of the mass parameters of the $\SU(N)$ flavor symmetry $\CS^\text{4d}\{\fg,\fg,\SU(N)\}$.

Now, let us use the class S technology to go to a different duality frame of the quiver \eqref{eq:qv1-4d} where the punctures are ordered as 
\begin{equation}
\sT_\fg\{Y_\text{simple},\ldots,Y_\text{simple},\g,\g\}.\label{eq:simplecollisions}
\end{equation}
In Sec.~\ref{sec:detailedS} below, we will determine  the resulting quiver for $\fg=A_{k-1}, D_k, E_6$ 
using the known data, and we will find that the outcome has the form, when $N$ is sufficiently large,
\begin{equation}
\text{ a 4d generalized quiver} - \fg - \sT_\fg\label{eq:anotherframe}
\end{equation}where the 4d quiver part on the left turns out to be 
exactly the $T^2$ reduction of the quiver theory of the 6d conformal matter of type $(\varnothing, \fg)$. 
This conformal matter of type $(\varnothing, \fg)$ is obtained from the results of \cite{DelZotto:2014hpa} as follows.
First we consider the 6d theory realized at the intersection of Ho\v rava-Witten E8-wall and ALE singularity of type $\g$
with some number of M5-branes in M-theory. 
From the lists of \cite{DelZotto:2014hpa}, it can be shown that this 6d SCFT has a subspace of the tensor branch in which
we get a configuration,
\beq
\begin{array}{ccccccccc}
	\varnothing & \su(1) &\g_2& \g_3 & \cdots&\g_i&\cdots & \g_{N-1} & [\g] \\
	1&  2& 2 & 2 &\cdots&2&\cdots & 2 &  \\
\end{array},
\eeq
where the $(-1)$-curve supports nothing, the leftmost $(-2)$-curve supports $\su(1)$ (which is Kodaira singular fiber of type $I_1$ or $II$),
and the rest of the gauge groups $\g_i$ ($i=2,3,\cdots$) and (generalized) matter content are determined in \cite{DelZotto:2014hpa}.
Now we regard the $(-1)$-curve to be a flavor brane with the empty flavor symmetry $\varnothing$ by taking this curve to be infinitely large.
By shrinking the remaining $(-2)$-curves, we get the conformal matter of type $(\varnothing, \fg)$.

Let us denote this 6d theory as $\CT^\text{6d}_{(\varnothing,\fg),N}$, where $N-1$ is the number of the $(-2)$-curves.
Its $T^2$ reduction is, from the general discussion in Sec.~\ref{sec:general},
given by a 4d theory $\CS^\text{4d}_{(\varnothing,\fg),N}\{\SU(N),\fg\}$ 
whose $\SU(N)$ flavor symmetry is gauged by an $\SU(N)$ multiplet 
with $\SL(2,\bZ)$ duality symmetry. We will prove in Sec.~\ref{sec:Coulomb} that this 4d theory $\CS^\text{4d}_{(\varnothing,\fg),N}\{\SU(N),\fg\}$ is an SCFT
without an IR free gauge group.
Furthermore, its $\SU(N)$ mass deformation gives the 4d generalized quiver appearing in \eqref{eq:anotherframe}.

Therefore, we conclude that the $T^2$ compactification of the theory $\CT^\text{6d}_{N}\{\fg,\fg\}$, i.e.~the theory on $N$ M5-branes probing the $\bC^2/\Gamma_{\fg}$ singularity, has the structure \begin{equation}
\frac{\CS^\text{4d}_{(\varnothing,\fg),N}\{\SU(N),\fg_T\} \times \sT_\fg\{\fg_B,\fg_L,\fg_R\}} {\SU(N)_\tau \times (\text{diag.~of  $\fg_T\times \fg_B$}) }
\end{equation} where $\SU(N)$ is conformal and the gauged diagonal part of $\fg_T\times \fg_B$ is infrared-free.\footnote{Note that we have $\fg_T=\fg_B=\fg_L=\fg_R=\fg$ here. The subscripts are there to distinguish various factors.}

Indeed, the theory $\CS^\text{4d}_{(\varnothing,\fg),N}\{\SU(N),\fg_T\}$ has the flavor central charge 
$k_{\fg_T} = 2h^\vee_{\fg_T} +2$ as shown in Sec.~\ref{sec:Coulomb},
and the theory $\sT_\fg\{\fg_B,\fg_L,\fg_R\}$ has the flavor central charge $k_{\fg_B} =2h^\vee_{\fg_B}$. Then the total flavor central charge of the two superconformal sectors for the gauged diagonal part is larger than $4h^\vee_{\fg}$ and the gauging is infrared-free. The flavor central charge $k_{\fg_T}$ for $\CS^\text{4d}_{(\varnothing,\fg),N}\{\SU(N),\fg_T\}$ is obtained from the formula \eqref{eq:baseAformula} and the fact that 
6d anomaly polynomial for the 6d theory $\CT^\text{6d}_{(\varnothing,\fg),N}$ contains $I \supset \frac{h^\vee_{\fg}}{24} p_1(T)c_2(\fg)$.

When $\fg=A_{k-1}$, the analysis in the previous section tells us that in fact \begin{equation}
\CS^\text{4d}_{(\varnothing,\fg),N}\{\SU(N),\fg\} = \sT_N\{[1^N],[1^N],[N-k,1^k]\},
\end{equation}
but we do not have a direct identification of $\CS^\text{4d}_{(\varnothing,\fg),N}\{\SU(N),\fg\}$ 
for $\fg\neq A$ 
with other known 4d SCFTs.

If one prefers a slightly more symmetric situation, one can start from the little string $\CT^\text{A}_{\fg,N}$ and its $T^2$ compactification is given by \begin{equation}
\frac{\CS^\text{4d}_{(\varnothing,\fg),N}\{\SU(N),\fg_T\} \times \sT_\fg\{\fg_B,\fg_L,\fg_R\}} {\SU(N)_\tau \times 
(\text{diag.~of $\fg_T\times\fg_B$})
 \times (\text{diag.~of $\fg_L\times\fg_R$})_{\tau'}},
\end{equation} which can also be obtained from the $T^2$ compactification of $\CT^\text{B}_{\fg,N}$.
In the compactification of $\CT^\text{A}_{\fg,N}$, the parameter $\tau$ is the complex structure of $T^2$, and the parameter
$\tau'$ is the complexified K\"ahler parameter of $T^2$. By the T-duality, the role of complex structure and complexified K\"ahler parameter are exchanged.

\subsection{Detailed class S analysis}\label{sec:detailedS}
Now what is left is to present a class S analysis  for the \eqref{eq:simplecollisions} for  $\fg=A_{k-1},$ $D_k$, and $E_6$.
\paragraph{When $\fg=A_{k-1}$,} the resulting quiver is \begin{equation}
\su(1)-\su(2)-\su(3)-\cdots-\su(k-1)-\su(k)-\su(k)-\cdots - \sT_k
\end{equation} where we have bifundamentals between neighboring groups and 
one additional fundamental at the leftmost $\su(k)$,
as by now well-known and originally derived in \cite{Gaiotto:2009we}.
This is indeed the $T^2$ reduction of the $(\varnothing,\su(k))$ matter,
see (6.5) of \cite{DelZotto:2014hpa}.

\paragraph{When $\fg=D_k$,}  the resulting quiver can be found by the data compiled in \cite{Chacaltana:2011ze}. We find \begin{equation}
\su(1)-\sp(1)-\fg_2-\fso(9)-\fso(11)-\cdots-\fso(2k-1)-\fso(2k)-\fso(2k)-\cdots-\sT_{D_k}
\end{equation}
where the matters are, from the left, \begin{itemize}
\item a half-hyper in the doublet, 
\item a half-hyper in $\mathbf{2}\otimes\mathbf{7}$,
\item the $E_8$ Minahan-Nemeschansky theory whose $\fg_2\times \fso(9)\subset \fg_2\times \ff_4 \subset \e_8$ is gauged,
\item the  $D_5$ generalized bifundamental $\sB_{D_5}$ whose $\fso(9)\times \fso(11)\subset \fso(20)$ symmetry is gauged, \ldots,
\item the $D_{k}$ generalized bifundamental $\sB_{D_k}$ whose $\fso(2k-1)\times \fso(2k)$ symmetry is gauged, etc. 
\end{itemize}
This is indeed the $T^2$ reduction of the $(\varnothing,\fso(2k))$ matter,
see the un-numbered equation at the top  of p.~34 of \cite{DelZotto:2014hpa}.
Note that the  theory $\sB_{D_k}=\sT_{D_k}\{\fso(2k),\fso(2k),Y_\text{simple}\}$ has an enhanced flavor symmetry $\fso(4k)$ compared to what is apparent in the class S description, and its subgroup $\fso (2k-1)\times \fso (2k+1)$ is gauged in this construction.

\paragraph{When $\fg=E_6$,} the resulting quiver can be found by the data compiled in \cite{Chacaltana:2014jba}: we find \begin{equation}
\su(1)-\sp(1)-\fg_2-\f_4-\e_6-\e_6 \cdots -\cdots-\sT_{E_6}\label{varnothinge6}
\end{equation} where the matters are, from the left, 
\begin{itemize}
\item a half-hyper in the doublet, 
\item a half-hyper in $\mathbf{2}\otimes\mathbf{7}$,
\item the $E_8$ Minahan-Nemeschansky theory whose $ \fg_2\times \ff_4 \subset \e_8$ is gauged,
\item the  $E_6$ generalized bifundamental $\sB_{E_6}$ whose $\f_4\times \e_6$ symmetry is gauged.
\end{itemize}
This is indeed the $T^2$ reduction of the $(\varnothing,\e_6)$ matter,
see (6.7) of \cite{DelZotto:2014hpa}.

\paragraph{When $\fg=E_7$ and $E_8$,} the class S data for $\fg=E_7$ and $E_8$ are not yet available. Nonetheless, we consider the agreement we found so far is convincing enough that this correspondence works for all $\fg$. This can also be considered as a prediction for the repeated collision of the simple punctures in the class S theory of type $E_7$ and $E_8$. 
From the structure of $(\varnothing,E_{n=7,8})$ conformal matters given in (6.8) and (6.9), our prediction is that the class S theories of type $E_{n=7,8}$ with multiple simple punctures and two full punctures have a duality frame of the form \begin{equation}
\su(1)-\sp(1)-\fg_2-\f_4-\e_n-\e_n \cdots -\cdots-\sT_{E_n}\label{varnothingen}
\end{equation} where the matters are, from the left,
\begin{itemize}
\item a half-hyper in the doublet, 
\item a half-hyper in $\mathbf{2}\otimes\mathbf{7}$,
\item the $E_8$ Minahan-Nemeschansky theory whose $ \fg_2\times \ff_4 \subset \e_8$ is gauged,
\item a certain SCFT with $F_4\times E_n$ flavor symmetry, which comes from the 6d very Higgsable theory with the structure \begin{equation}
\begin{array}[b]{cccccccc}
[\mathfrak{f}_4]&&\mathfrak{g}_2&\mathfrak{su}_2&&[\mathfrak{e}_7]\\ \relax
&1&3&2&1&
\end{array}\quad \text{for $E_7$},\qquad
\begin{array}[b]{cccccccc}
[\mathfrak{f}_4]&&\mathfrak{g}_2&\mathfrak{sp}_1&&&[\mathfrak{e}_8]\\ \relax
&1&3&2&2&1&
\end{array}\quad \text{for $E_8$},
\end{equation}
\item and the $E_n$ generalized bifundamentals $\sB_{E_n}$ which is the class S theory on  a sphere with two full punctures and a simple puncture.
\end{itemize}

\section{The 4d tensor branch and IR free gauge group}\label{sec:Coulomb}
In this section, we discuss more details of the structure of the branch $\CCt$ discussed in Sec.~\ref{sec:general}.
This is the branch where only Coulomb moduli coming from the tensor multiplets associated to ($-2$)-curves get vevs.
Our discussion in this section concentrates on four dimensions.
We also give general arguments about the existence of an IR free gauge group and a sufficient condition under which they are absent.

\subsection{The 4d tensor branch $\CCt$}
Let us recall and extend what we have found in Sec.~\ref{sec:general}.
The 4d Coulomb moduli subspace $\CCt$ is the same as that of the compactification of \Nequals{(2,0)} theory of type $G$ on a torus with nonzero area.
The compactification of \Nequals{(2,0)} theory on $S^1$ has a subtle discrete parameter that controls the global structure and the discrete theta angle of the gauge group $G$, see e.g.~\cite{Witten:2009at,Tachikawa:2013hya,DelZotto:2015isa} and this choice also affects the moduli space. 
Our aim is to study the behavior at the most singular point on the moduli space, and it does not depend on the choice of this discrete parameter.

The 4d tensor branch $\CCt$ is given by a finite quotient of  $(\bR \times S^1)^{\rank G}$.
In terms of the 6d tensor multiplets $(\phi_i, B_i)$,
the 4d variables are defined as
\beq
u_i = \int_{T^2} ( B_i + \ii \phi_i ).
\eeq
The periodicity depends on the precise choice of the discrete parameter. 
One choice is to take $u_i \sim u_i+1$.
This gives the moduli space of the 5d gauge theory on $S^1$ where the gauge group is simply connected.
Another choice is to take $u^i \sim u^i+1$ where $u^i=\eta^{ij} u_i$. 
This gives the moduli space of the 5d gauge theory on $S^1$ where the gauge group is the adjoint group.
The condition of positive volumes of ($-2$)-curves is $\phi^i \geq 0$, which
corresponds to a Weyl chamber in $\bR^{\rank G}/W$ where $W$ is the Weyl group.

Generically, each ($-2$)-curve supports a non-Abelian gauge group whose field strength is denoted as $F_i$.
When ${\rm Im}( u^i) \to \infty$,
the effective action of these gauge groups in 4d with Euclidean signature is given by
\beq
-S_E \to \int   \left(  2 \pi \ii u^i \cdot \frac{1}{4}  \Tr (F_i^- \wedge F_i^- ) + 2 \pi \ii (u^i)^*  \frac{1}{4}  \Tr (F_i^+ \wedge F_i^+ )   \right) 
\label{eq:weakeff}
\eeq
where $F_i^\pm = (F_i  \pm \star F_i)/2$, and the normalizations of $\Tr$ and $F_i$ are such that one instanton gives 
$\int \frac{1}{4}   \Tr (F_i^- \wedge F_i^- ) =1$ and $F_i^+=0$. This action can be obtained by the dimensional reduction of \eqref{eq:eff5d} to 4d.
Note that the nonrenormalization theorem of Sec.~\ref{sec:general} applies only to the kinetic terms of $u^i$.
The kinetic terms of $F_i$ are subject to quantum corrections. Thus the above expression holds only in the weak 
coupling limit ${\rm Im}( u^i) \to \infty$. This is the limit where the tensor multiplet vevs $\phi^i$ are much larger than the compactification scale of $T^2$.

\subsection{$\CCt$ as a space of exactly marginal deformation}\label{sec:marginal}

On the generic point of $\CCt$, we have a quiver gauge theory of the gauge groups on ($-2$)-curves
with very Higgsable generalized matters. 
We showed in Sec.~\ref{3.4} that this quiver theory is superconformal.
Therefore, there is a space 
of exactly marginal deformations of this quiver theory which we denote as $\CM_{\rm marginal}$.
The gauge couplings are determined by the vevs of the moduli fields $u^i$ as described above.
Then, we have a map
\beq
\pi: \CCt \to \CM_{\rm marginal}.
\eeq
If none of the gauge group supported on the curve is  empty (i.e.~$\su(1)$), the dimension of $\CM_{\rm marginal}$ is given by $\rank G$, which is the same as the dimension of $\CCt$. 
Then the existence of the above map implies that $\CCt=(\bR \times S^1)^{\rank G}/W$ is a finite covering of $\CM_{\rm marginal}$.\footnote{
Infinite covering is impossible because of the explicit form of $\pi$ in the weak coupling limit ${\rm Im}(u^i) \to \infty$ determined by \eqref{eq:weakeff}.}

To make the argument more explicit, let us focus on the case $G=A_{N-1}$. Then we can describe $\CCt$ in the following way.
First, we introduce variables $v_i$ ($i=1,\cdots,N$) such that
\beq
\diag(v_1, v_2,\cdots,v_N)= \sum_i u_i H^i,
\eeq
where $H^i=(0,\cdots,0,1,-1,0,\cdots,0)$ are the Cartan generators of $\SU(N)$.
Furthermore, instead of imposing a traceless condition, we impose an equivalence relation
\beq
(v_1,\cdots,v_N) \sim (v_1,\cdots,v_N)+\alpha(1,\cdots,1)
\eeq
for arbitrary $\alpha$. This means that we are considering the 5d group as $G ={\rm PSU}(N)=\SU(N)/\bZ_N$.

We introduce $\bC^\times \cong \CP^1 \setminus \{0, \infty\}$ variables as $z_i=e^{2\pi \ii  v_i}$.
The equivalence relation is now
\beq
(z_1,\cdots,z_N) \sim c(z_1,\cdots,z_N)
\eeq
for $c \in \bC^\times$. The Weyl group $W$ is the symmetric group ${\mathfrak S}_N$ which permutes the variables $z_i$.

Now we can see that the space $\CCt$ for $G=\SU(N)/\bZ_N$ is the same as the complex structure moduli space of
a Riemann sphere $\CP^1$ with two distinguished points $0$ and $\infty$, and $N$ indistinguishable points $z_i $.
We call the points $0$ and $\infty$ as punctures $Y_L$ and $Y_R$, and the points $z_i$ as simple punctures $p_i$.
We may call this moduli space as $\CM_{\CP^1} \{Y_L,Y_R, p_1,\ldots,p_N \}$.

When the quiver is a class S theory as in the examples in Sec.~\ref{sec:examples1} and \ref{sec:examples2}, 
the above moduli space $\CM_{\CP^1} \{Y_L,Y_R, p_1,\ldots,p_N \}$ 
is exactly as expected for the moduli space $\CM_{\rm marginal}$ of the exactly marginal deformations of the quiver theory.
So we expect that the map $\pi: \CCt=\CM_{\CP^1} \{Y_L,Y_R, p_1,\ldots,p_N \} \to \CM_{\rm marginal}$ is the natural one. One can explicitly check that this is the case
in the weak coupling limit by using \eqref{eq:weakeff} and the relation $2\pi \ii u^i = \log(z_i/z_{i+1})$.
When $Y_L$ and $Y_R$ in the class S are different,
the map $\pi$ is just isomorphism. When $Y_L$ and $Y_R$ are identical in the class S, the map $\pi$
is 2 to 1 because we have to divide by the $\bZ_2$ symmetry $Y_L \leftrightarrow Y_R$ in  $\CM_{\rm marginal}$.
If the group $G$ is $\SU(N)$ instead of $\SU(N)/\bZ_N$, the moduli space $\CCt$ is given by the $N$-covering of 
$\CM_{\CP^1} \{Y_L,Y_R, p_1,\ldots,p_N \}$.

It needs to be  stressed that our argument above, that is the existence of $\pi: \CCt \to \CM_{\rm marginal}$, 
did not require that the quiver is class S.
We can consider more general theories with very Higgsable matters. Even in that case, $\CM_{\rm marginal}$ is given by the moduli of a Riemann sphere
with $N+2$ punctures $\{Y_L,Y_R, p_1,\ldots,p_N\}$ up to a division by some discrete group.
A simple example is the theory of a single ($-2$)-curve supporting an $E_8$ gauge group which is coupled to a rank 10 E-string theory,
which has no known class S description to the authors' knowledge.%\note{Is there really no class S description of this theory?}

\subsection{Singular loci and IR free gauge group}\label{sec:SingIRfree}
Now we consider singular loci of $\CCt$. On generic points of $\CCt$, the group $G$ is broken down to $\U(1)^{\rank G}$.
But there are singular loci $u^i =0$ where a subgroup of $G$ is recovered and new massless degrees of freedom appear.

Let us again consider the case $G=\SU(N)/\bZ_N$. These singular loci are sub-varieties of $\CCt$ given by $z_i=z_j$ for some $i$ and $j$.
These are exactly the loci where simple punctures $p_i$ and $p_j$ collide on the Riemann sphere. 
In particular, the most singular point is given by $z_1=z_2=\cdots=z_N$, where $N$ simple punctures collide.
In terms of the Riemann sphere with $N+2$ punctures, this is a degeneration limit in which
the sphere is decoupled into a sphere with three punctures $\{Y_L, Y_R, Y\}$ and another sphere with $N+1$ punctures $\{Y, p_1,\ldots, p_N\}$
for a puncture $Y$. In this degeneration limit, we may expect that there is a gauge coupling in some duality frame 
which becomes infinitely weak. 
This weak gauge group $H_{\rm weak}$ is coupled to the puncture $Y$ on both sides, at least in the case of class S. 
This is very likely to be the case for more general theories as in Argyres-Seiberg-Gaiotto dualities \cite{Argyres:2007cn,Gaiotto:2009we}.
Other gauge couplings are determined by the ratio of the vevs of $u^i$ in the limit $u^i \to 0$.

Combined with what has been discussed repeatedly in this paper, we get the following picture of the above limit.
The theory with $N+1$ punctures $\{Y, p_1,\ldots,p_N\}$ is actually obtained as a low energy effective theory of some theory $\cU\{G,Y\}$ 
with a flavor $G$ symmetry which is coupled to the $G$ gauge field. 
The vevs of $u^i$ give mass deformation of the theory $\cU\{G,Y\}$, and after the mass deformation we get 
the low energy theory corresponding to the sphere with $N+1$ punctures.
The gauge group $H_{\rm weak}$ which is infinitely weakly coupled in the limit $u^i \to 0$ is actually an IR free gauge group before
the vevs of $u^i$ are turned on. That is, the contribution of the $\cU\{G,Y\}$ to the beta function of $H_{\rm weak}$ is such that
this gauge group is IR free, and after the mass deformation, the contribution of the theory to the beta function is changed 
and $H_{\rm weak}$ becomes conformal below the mass scale determined by a certain average of the vevs of $u^i$.
Therefore the coupling of $H_{\rm weak}$ is roughly given as 
\beq
\tau_H \sim \frac{b_0}{2 \pi \ii }  \log (\overline{u}),\label{eq:Hcoup}
\eeq
where $b_0$ is the coefficient of the IR free beta function, and $\overline{u}$ is a certain average of $u^i$.
Note that the log term may be written more physically as $\log (\overline{u}\overline{R}^{-1}/ \overline{R}^{-1})$,
where $\overline{R}$ is the scale of compactification of $T^2$, and $u^i \overline{R}^{-1}$ are fields with canonically normalized kinetic terms
whose vevs give the mass scale of the mass deformation.
In this form, it is clear that the Landau pole of the IR free gauge group is located at the compactification scale $\overline{R}^{-1}$.

The running of the coupling \eqref{eq:Hcoup} explains the fact
that $H_{\rm weak}$ becomes infinitely weakly coupled in the degeneration limit $u^i \to 0$ of the Riemann sphere with $N+2$
punctures $\{Y_L,Y_R,p_1,\ldots,p_N\}$. Let us also denote the theory corresponding to the Riemann sphere with three punctures $\{Y_L, Y_R, Y\}$
as $\cV\{Y_L,Y_R,Y\}$. We conclude that the theory at the most singular point is given by
\beq
\frac{ \cU\{G,Y\}  \times \cV\{Y_L,Y_R,Y\}  }{ H_{\rm weak} \times G_\tau}, \label{eq:genIRfree}
\eeq
where $H_{\rm weak}$ gauges the punctures $Y$.
The theory without gauging $G$ is the one we have denoted $\CS^{4d}\{G\}$ in Sec.~\ref{sec:general}.

The above arguments are the abstract version of what we have discussed in Sec.~\ref{sec:simplest},  \ref{sec:examples1} and \ref{sec:examples2}.
For example, in the case of A-type conformal matters with deformation discussed in Sec.~\ref{sec:GT}, 
we have $\cU\{G,Y\}/G=\sS_N \vev{T^2_\tau}\{ Y  \}$ and $\cV\{Y_L,Y_R,Y\} =\sT_k\{ Y_L, Y_R,Y\}$
where $Y$ is of the form $Y=[*, 1^{{\rm min}(N,k)}]$.

We again emphasize that the above picture is expected to be correct even in theories without class S realizations.
The only assumption which we do not completely prove is that the degeneration limit in the moduli space $M_{\rm marginal}=\pi(\CCt)$ discussed
above implies the appearance of an infinitely weakly coupled gauge group $H_{\rm weak}$ in the conformal quiver, which is very likely to be the case.
We summarize what we have found when $G=A_{N-1}$. We expect that  things work out similarly for more general $G$.
\begin{claim}
1. The quiver realized on ($-2$)-curves is a conformal quiver in 4d with very Higgsable generalized matters.
When the configuration of ($-2$)-curves is of $G=A_{N-1}$ type, the space of exactly marginal deformation $\CM_{\rm marginal}$ 
is given by a complex structure moduli space of a Riemann sphere with two distinguished points $Y_L, Y_R$ and $N$ indistinguishable points $p_i$,
possibly divided by some discrete group. They show generalized Argyres-Seiberg-Gaiotto dualities.

\bigskip

\noindent 2. At the most singular point of the 4d tensor branch $\CCt$, the $G$ gauge group is restored whose
gauge coupling $\tau$ is the complex structure modulus of $T^2$ with $\SL(2,\bZ)$ action. At this point, we get two theories
connected by an IR free gauge group $H_{\rm weak}$ as in \eqref{eq:genIRfree}, if the gauge groups on ($-2$)-curves are  generic. 
By giving vevs to the vector multiplet of $G$,
a part of the theory is mass deformed and we get a low energy theory which is the above conformal theory in a certain duality frame.
\end{claim}

There is a situation in which the IR free gauge group $H_{\rm weak}$ disappears. A sufficient condition for the absence of the IR free gauge group is
the following. Suppose that one of the gauge groups on the $(-2)$-curves is
trivial, i.e., it is $\su(1)$.\footnote{In F-theory language, this $\su(1)$ is a singular fiber of Kodaira type $I_1$ or $II$.} 
The classification in \cite{Heckman:2015bfa} shows that $\su(1)$ can appear only at the ends of the quiver.
Suppose that there is one $\su(1)$. Then the dimension of $\CM_{\rm marginal}$ is $\rank G-1$, while that of $\CCt$
is still $\rank G$. In this case, $\pi: \CCt \to \CM_{\rm marginal}$ is not a finite covering, but a map from $\rank G$ dimensional space to $\rank G-1$
dimensional space.

The experience in Sec.~\ref{sec:examples1} and \ref{sec:examples2} of the class S case suggests the following interpretation of this situation.
In this case, we discard one of the punctures $Y_L$ or $Y_R$, say $Y_L$. Then, $\CM_{\rm marginal}$ is given
by a moduli space of a Riemann sphere with $N+1$ punctures $\{Y_R, p_1,\ldots,p_N \}$, divided by some discrete group.

Let us collide the punctures $p_i$ in this situation by taking the limit $z_1=\cdots =z_N$ or equivalently $u^i=0$. 
Then, one of the Riemann spheres after degeneration has only two punctures $\{Y_R,Y\}$; 
this is actually not a limit in which a Riemann sphere degenerates into two.
Then we cannot use the logic which have led us to the IR free gauge group in the above discussion. 
Indeed, in the following subsections we will show that there is no IR free gauge group in this situation by direct computation
similar to the one performed in \cite{Ohmori:2015pua}.

\subsection{Generalized blow-downs, intersection numbers and anomalies}\label{sec:self}
Before doing the computation involving an $\su(1)$ gauge group, we need some preparation. 
The discussions in this subsection might have broader applications than the cases considered in this paper.
The results here are likely to be related to those in \cite{DelZotto:2014fia}, but we do not go into details on this point.

When we have a configuration of curves with self intersection numbers $-x$, $-1$, and $-y$ as
$
x,1,y,
$
then by blowing down the $-1$ curve, we get
\beq
x,1,y \xrightarrow{\text{blowdown}} (x-1),(y-1).
\eeq
We want to generalize this to the case of blowing down a curve of arbitrary self-intersection number $-z$ in a configuration like
$
x, z, y.
$
After blowing down the curve $-z$, we get a singular base. However, even in such a singular base, the
intersection numbers have a clear field theoretical meaning as follows.
 
The intersection matrix $\eta^{ij}=-C^i \cdot C^j$ of curves $C^i$
gives the kinetic term of tensor multiplets $\phi_i$,
\beq
\eta^{ij} \partial_\mu \phi_i \partial^\mu \phi_j.
\eeq
The size of the curve $C^i$ is given by the scalar vevs of the tensor multiplets as $\phi^i = \eta^{ij} \phi_j$.
Now let us blow down a curve $C^k$ for some $k$. Then, we get a system of a strongly interacting SCFT involving $\phi^k$ 
and free tensor multiplets $\phi_i $ with $i \neq k$.
The above kinetic term is changed as 
\beq
\sum_{i,j \neq k }\eta'^{ij} \partial_\mu \phi_i \partial^\mu \phi_j+(\text{strongly interacting part}).
\eeq
The new matrix $\eta'^{ij}$ can be obtained by setting the size of the cycle $C^k$ to be zero;
\beq
\phi^k=\eta^{kk} \phi_k + \sum_{j \in K} \eta^{kj} \phi_j \to 0,
\eeq
where $j\in K$  means that  $C^j$  intersects with $C^k$.
Imposing this constraint, the kinetic term $\eta^{ij} \partial_\mu \phi_i \partial^\mu \phi_j$ becomes,
\beq
 \eta^{ij} \partial_\mu \phi_i \partial^\mu \phi_j
= \sum_{i,j \neq k} \eta^{ij} \partial_\mu \phi_i \partial^\mu  \phi_j - \sum_{i \in K, j \in K} \frac{\eta^{ik}\eta^{jk}}{\eta^{kk}} \partial_\mu \phi_i \partial^\mu \phi_j
+\frac{1}{\eta^{kk}} \partial_\mu \phi^k \partial^\mu \phi^k.
\eeq
The last term is incorporated into the strongly interacting part. So we get
\beq
\eta'^{ij}=\eta^{ij} - \frac{\eta^{ik}\eta^{jk}}{\eta^{kk}}.
\eeq
We call this  $\eta'^{ij}$  the generalized intersection number after the blow-down.
 
Now consider a configuration 
\beq
X_L,x_1,x_2, \cdots,x_r ,X_R \label{eq:genA}
\eeq
where $X_L$ and $X_R$ are non-compact cycles.
If we blow-down the leftmost curve of self-intersection $-x_1$, we get
\beq
X_L,\vev{\frac{1}{x_1}},x_2-\frac{1}{x_1},x_3,\cdots, x_{r},X_R
\eeq
where the number in the angle bracket $\vev{\cdot}$ is the intersection number of the adjacent curves $X_L$ and $x_2-\frac{1}{x_1}$. 
If we do not write this angle bracket, that means the intersection is $1$ as usual.
In other words we regard the angle bracket $\vev{1}$ as implicit.
Continuing the blow-down, we get a single curve
\beq
X_L,\vev{y_r},z_r , X_R
\eeq
where $z_r$ and $y_r$ are determined recursively by
\beq
z_r=x_{r}-\frac{1}{z_{r-1} },~(z_1=x_1);~~~~~y_r = \frac{y_{r-1}}{z_{r-1}},~(y_1=1).
\eeq
In particular, when all the curves are $x_i=2$, we get 
\beq
z_r=\frac{r+1}{r},~~~y_r=\frac{1}{r}
\eeq

Next, let us consider the effect of the blow-downs to the effective action.
Suppose that the tensor multiplet two-forms $B_i$ have the term
$
 2\pi \ii B_i \wedge I^{i}.
$
Rewriting this as
\beq
 B_i \wedge I^{i} =\frac{1}{\eta^{kk}}(\eta^{kk}B_k +\sum_{j \in K} \eta^{k j} B_j  )I^k+ \sum_{j \neq k} B_j \wedge (I^j-\frac{\eta^{jk}}{\eta^{kk}} I^k),
\eeq
we get the term after blowdown of $C^k$ as
\beq
\sum_{i \neq k} B_i \wedge I'^i, ~~~~
I'^i = I^i-\frac{\eta^{ik}}{\eta^{kk}} I^k.
\eeq
The $I^i$ has a form \cite{Ohmori:2014kda}
\beq
I^i \supset  d^i c_2(R)+\eta^{ij} c_2(F_j), \label{eq:c2form}
\eeq
where $c_2(R)$ is the second Chern class of the $\SU(2)_R$ R-symmetry, and $c_2(F_i)=\frac{1}{4} \Tr F_i^2$ are the second Chern classes
of gauge or flavor fields. We omitted a term containing $p_1(T)$ because it will not be used  in our discussions below.
Then, after the blow-down of $C^k$, we get 
\begin{equation}
\begin{aligned}
I'^i &\supset d^i c_2(R)   +   \eta^{ij} c_2(F_j) -\frac{\eta^{ik}}{ \eta^{kk}} (d^k c_2(R)+\eta^{jk} c_2(F_j)) \\
&=d'^ic_2(R)+\eta'^{ij}c_2(F_j) 
\end{aligned}
\end{equation}
where $d'^i=d^i- \frac{\eta^{ik}}{\eta^{kk}} d^k$.
Note that the coefficient of the terms $c_2(F_i)$ is still given by  the new intersection matrix $\eta'^{ij}$ after the blow-down.

In particular, consider the configuration \eqref{eq:genA}. 
If the curve $x_i$ in \eqref{eq:genA} has the term $d^i c_2(R)$, then after blowing down $r-1$ times, we get a recursion relation
\beq
d'^{r}= d^{r}+ \frac{ d'^{r-1}  }{  z_{r-1}  }. 
\eeq
When $x_i=2$ for all $i$, we get
\beq
d'^r=\frac{1}{r} \sum_{i=1}^{r} i d^i.\label{eq:dprime}
\eeq

\subsection{Theories without IR free gauge group, type $A$}
Here we show that we get a 4d SCFT without an IR free gauge group if the gauge group at one end of the quiver is $\su(1)$.
In this subsection we focus on the case $G=A_{N-1}$.
We use the moduli space when  the 5d gauge group is $G=\SU(N)/\bZ_N$,  just for simplicity of computation, because in this case there is only one point on $\CCt$ at which the gauge symmetry $G$ is restored.
The final result does not depend on the choice of the center of $G$.
As a byproduct, we will also see that the consistency of the analysis requires that an IR free gauge group is absolutely necessary  without $\su(1)$, when the configuration of $(-2)$ curves is type $A$.

What we will show is the following. We consider a configuration
\beq
\begin{array}{ccccccccc}
[\varnothing]	& \su(1) &\g_2& \cdots&\g_i&\cdots & \g_{N-1} &  [\g_R]\\
	X_L&  2& 2 &\cdots&2&\cdots&2& X_R\\
	   &&&&X_i\\
	   &&&&[\f_i]
\end{array}, \label{eq:su1theory}
\eeq
where the leftmost gauge group is $\g_1=\su(1)$, and $X_L$, $X_i$ and $X_R$ are non-compact cycles supporting flavor symmetries $\varnothing,\f_i$, and $\g_R$, respectively.
There are $N-1$ types of flavor branes $X_i~(i=1,2,\cdots,N-1)$, and $X_L$ and $X_R$ are special cases of $X_i$ ($i=1$ or $N-1$), 
but we explicitly write $X_L$ and $X_R$ since they play an important role in the proof. 
Let $\CT^\text{6d,4d} (\g_1,\cdots,\g_{N-1}[\g_R])$  denote the theory at the most singular point of this tensor branch structure and its $T^2$ compactified version.
This class of theories includes, but not restricted to, the theories which reduce to $\CS^\text{4d}_{(\varnothing,\fg),N}\{\SU(N),\fg\}/\SU(N)$ in 4d
discussed in Sec.~\ref{sec:notdetailedS} and \ref{sec:detailedS}.
We will prove:
\begin{claim} 
1. $\CT^\text{4d}(\g_1,\cdots,\g_{N-1}[\g_R])$ is conformal if $\g_1=\su(1)$, and

\bigskip

\noindent 2. the effective numbers of hyper and vector multiplet $n_h,n_v$, and the flavor central charges 
$k_{\f_i},k_{\g_{R}}$ of $\CT^\text{4d}(\su(1),\cdots,\g_{N-1}[\g_{R}])$ are given by
\begin{equation}
\begin{aligned}
	 n_h-n_v&=q, &	n_v&=2r-p+2\sum_{i=1}^{N-1}(N-i)d^i,\\
	k_{\f_i}&=2s_{\f_i}+2(N-i),& k_R&=2s_R+2,
\end{aligned}
	\label{eq:baseAformula}
\end{equation}
where $p,q,r$, $d^i$ and $s_{\f_i},s_R$ are the coefficients of the 6d anomaly polynomial $I_\text{matter}$ before adding 
Green-Schwarz contribution from tensor multiplets of $(-2)$-curves as 
% $\CT^\text{6d}(\g_1,\cdots,\g_{N-1}[\g_R])$ defined by
\begin{multline}
	I_\text{matter} \supset  p \frac{c_2(R)p_1(T)}{48}+q \frac{7 p_1(T)^2 - 4 p_2(T)}{5760} + r \frac{p_1(T)^2-4p_2(T)}{192} \\
	- d^i c_2(F_{\g_i}) c_2(R) + s_{\f_i} \frac{p_1(T) c_2(F_{\f_i})}{24}+s_{R} \frac{p_1(T) c_2(F_{R})}{24}.\label{eq:6danm}
\end{multline}
The $d^i$ are also found from the couplings of two form fields $B_i$ of the tensor multiplets 
$2 \pi \ii B_i I^i$, where $I^i=d^i c_2(R)+\eta^{ij}c_2(F_{\g_i})- c_2(F_{\f_i})$.
\end{claim} 
The Green-Schwarz contribution of $(-2)$-curves is given by $I_\text{GS}=\frac{1}{2} \eta_{ij}I^i I^j$,
and this contribution plays the important role in gauge anomaly cancellation~\cite{Sadov:1996zm} and 
flavor anomaly matching~\cite{Intriligator:2014eaa,Ohmori:2014kda}. (See \cite{Ohmori:2014kda} for the conventions used in this paper.)
In the total anomaly $I_\text{total}=I_\text{matter}+I_\text{GS}$, 
the terms containing gauge fields $c_2(F_{\g_i}) $ are cancelled, the term proportional to $p_2(T)$ does not change, and
the terms containing $p_1(T)$ do not change in the case of $(-2)$-curves. 
Therefore the coefficients $p,q,r$ and $s_{\f_i},s_R$ are the same both in $I_\text{matter}$ and $I_\text{total}$,
and the term $-d^i c_2(F_{\g_i}) c_2(R)$ is absent in $I_\text{total}$.

Note that the discussion in Sec.~\ref{sec:SingIRfree} implies that the SCFT $\CT^\text{4d}(\g_1,\cdots,\g_{N-1}[\g_R])$ includes a conformal gauge group 
$\SU(N)$ and the $\SL(2,\bZ)$ modular transformation of the compactifying torus acts on the conformal coupling of the gauge field as the S-duality operation.

\subsubsection{Outline of the proof}
The proof will be done  inductively,  similarly to what was done for the very Higgsable theories argued in \cite{Ohmori:2015pua}.
We blow down  $N-2$ curves from the left using the results of Sec.~\ref{sec:self}, and find
\beq
\begin{array}{ccccccccc}
	& & \g_{N-1} & [\g_R] \\
X_L& \vev{\frac{1}{N-1}} &\frac{N}{N-1}& X_R\\
	  &&\vev{\frac{i}{N-1}}\\
	  &&X_i\\
	  &&[\f_i]
\end{array}.\label{eq:1dsub}
\eeq
This configuration gives the theory $\CT^\text{6d}(\g_1,\cdots,\g_{N-2}[\g_{N-1}])/\g_{N-1}$ plus a tensor multiplet and 
4d very Higgsable matters with the symmetry  $(\g_{N-1} , \g_R)$ which are coupled to the $\g_{N-1}$ gauge field.
At this  step of the induction, we assume that 
\begin{itemize}
		\itshape
	\item \textsc{Assumption$_{N-1}$:} $\CT^\text{4d}(\g_1,\cdots\g_{N-2}[\g_{N-1}])$ is conformal and the flavor central charge $k_{\g_{N-1}}$ of $\g_{N-1}$ is $k_{0,\g_{N-1}}+2$ where $k_{0,\g_{N-1}}$ is the flavor central charge of the very Higgsable matters with the symmetry $(\g_{N-2},\g_{N-1})$.
		Equivalently, the gauge group $\g_{N-1}$ in the configuration \eqref{eq:1dsub} is
IR free, and the beta function of this IR free gauge group is $2$ in the normalization that the gauge multiplet contribution is $-4h^\vee$,
where $h^\vee$ is the dual Coxeter number of the gauge group. 
\end{itemize}
Then we will prove that 
\begin{itemize}
		\itshape
	\item at the most singular point of the Coulomb branch of $\CT^\text{4d}(\g_1,\cdots,\g_{N-1}[\g_R])$, the flavor central charges of $\fg_{N-1}$ and $\g_R$ get an additional contribution
		\begin{equation}
	       	\delta k_{\fg_{N-1}}=-2, \qquad \delta k_{\fg_R}=2,
		\end{equation}
		therefore \textsc{Assumption$_{N}$} holds, and 
	\item at the most singular point of the Coulomb branch of $\CT^\text{4d}(\g_1,\cdots,\g_{N-1}[\g_R])$, the conformal central charges $n_v,n_h$ and the flavor central charge $k_{\f_i}$ for $\f_i$ get additional contribution 
		\begin{equation}
			\delta n_v=\delta n_h=2\sum_{i=1}^{N-1}d^i, \qquad \delta k_{\f_i}=2.
		\end{equation}
\end{itemize}
Accumulating the $\delta n_v,\delta n_h, \delta k_{\f_i}$, we will get the result \eqref{eq:baseAformula}.

\subsubsection{Method}
We use the method developed in \cite{Shapere:2008zf} which was also used in a similar context when we studied 
very Higgsable theories in \cite{Ohmori:2015pua}.
Suppose that we have a Coulomb moduli (sub)space $\CC$ which we assume to be one-dimensional in this paper.
We take the coordinate as $w $. Then, on this Coulomb branch, we have an effective action 
including background fields as \cite{Witten:1995gf}
\beq
-S_E \supset \int (\log D(w)) c_2(R)+(\log E(w)) p_1(T)+(\log C(w)) c_2(F).
\eeq
The $D(w)$, $E(w)$ and $C(w)$ are holomorphic by supersymmetry.\footnote{Strictly speaking, we need to do Donaldson-Witten twisting for this statement
to be true. But this point is not important in our discussion below.}
Here $(\log C(w)) c_2(F)$ must be replaced by a sum $\sum_k (\log C_k(w)) c_2(F_k)$ if we have
multiple gauge and flavor symmetries, but we just write this as $(\log C(w)) c_2(F)$ for notational simplicity.

There are some points $w_a$ on $\CC$ at which new massless degrees of freedom appear. Near such points,
the low energy effective action can become singular. 
Then locally near $w_a$, we have
\beq
D(w) \sim (w-w_a)^{\alpha_a},~~~E(w) \sim (w-w_a)^{\beta_a},~~~C(w) \sim (w-w_a)^{\gamma_a}.
\eeq
If we go around the singular point, we get a monodromy 
\beq
\log D(w) &\to \log D(w) +2 \pi \ii \alpha_a, \nonumber \\
\log E(w) &\to \log E(w) + 2\pi \ii \beta_a, \nonumber \\
\log C(w) &\to \log C(w) + 2 \pi \ii \gamma_a.
\eeq
This monodromy is due to the massless degrees of freedom at the singular point $w_a$. This is a familiar phenomenon in field theory,
especially in Seiberg-Witten solutions \cite{Seiberg:1994rs} near massless monopole points.

Now we may define some $\U(1)_R$ symmetry near the singular point. This $\U(1)_R$ does or does not have gauge anomalies
depending on whether the theory at the singular point is non-SCFT or SCFT, respectively. 
This $\U(1)_R$ is defined only locally near $w_a$ and need not be defined globally on the entire Coulomb space $\CC$.
Suppose that $(w-w_a)$ has $\U(1)_R$ charge $q_a$. Then,
the above monodromy contributes to the anomaly of the $\U(1)_R$ symmetry under the background and gauge fields as
\beq
\U(1)_R~\text{anomaly from monodromy}:  q_a ( \alpha_a c_2(R)+\beta_a p_1(T) +  \gamma_{a} c_2(F)). \label{eq:mon}
\eeq
This must be the anomaly due to additional massless degrees of freedom at the singularity. This is the familiar anomaly matching
involving Nambu-Goldstone bosons of spontaneously broken symmetries.

In general, the anomaly of the $\U(1)_R$ symmetry of a theory under background and gauge fields is given by
\beq
-n_v c_2(R) -\frac{n_v-n_h}{12}p_1(T)+k c_2(F)
\eeq
where $n_v$ and $n_h$ are effective numbers of vector multiplets and hypermultiplets which are defined in terms of central charges $a$ and $c$
as $a=\frac{5}{24} n_v+\frac{1}{24} n_h$ and $c=\frac{1}{6}n_v+\frac{1}{12} n_h$, and $k$ is the flavor central charge if the group is a flavor symmetry,
or the coefficient of the beta function if the group is a gauge symmetry.
The anomaly \eqref{eq:mon} accounts for the difference of this anomaly between the theory at the singular point and the theory at the generic point,
$\delta n_v=-q_a \alpha_a,~\delta (n_h-n_v)=12 q_a \beta_a$ and $\delta k = q_a \gamma_a$.

In general, linear combinations of the functions $D(w)$, $E(w)$
and $C(w)$ can have a very nontrivial nonabelian monodromy on $\CC$. In our context, the most crucial point is the following. 
If we have a gauge field $F_i$ which is conformal at generic points of $\CC$, then the function $C_i(w)$ corresponding to this
gauge field is essentially $e^{2 \pi \ii  \tau_i}$, where $\tau_i$ is the gauge coupling of $F_i$. 
In this case, this function $C_i(w)$ has a nonabelian monodromy
as discussed explicitly in Sec.~\ref{sec:simplest} and rather abstractly in Sec.~\ref{sec:marginal} due to S-duality of the gauge group.
However, if the gauge group is IR free on generic points of $\CC$, the $C_i(w)$ only has an abelian monodromy which is just a phase shift,
because an IR free gauge group does not have S-duality and there is only $T$ transformations $\tau_i \to \tau_i+\bZ$.

We will show below that there is only an IR free gauge group which is $\g_{N-1}$ on generic points of the one-dimensional subspace \eqref{eq:1dsub}.
Then, we will show that this IR free gauge group becomes conformal at the singular point $w_a$.
On the other hand, if we have a conformal group at generic points of $\CC$, then we get an IR free gauge group at the singularity $w_a$.
This is what was found in Sec.~\ref{sec:simplest} and Sec.~\ref{sec:SingIRfree} when the gauge groups $\g_i$ are generic without $\su(1)$.

We still need to worry about the nonabelian monodromy of $D(w)$ and $E(w)$. However, 
fortunately, there is a simplification in our situation. The space $\CC$ will be a one-dimensional subspace 
\eqref{eq:1dsub} of $\CCt$.
As we discussed, $\CCt$ is the same as that of the \Nequals{(2,0)} theory compactified on $T^2$. Due to the maximal supersymmetry of
the \Nequals{(2,0)} theory, the monodromy of $D(w)$ and $E(w)$ is only abelian, i.e, there is only a phase shift.
But we remark that the formalism of \cite{Shapere:2008zf}  can be applied even if we encounter nonabelian monodromy.

\subsubsection{Warm up: a single $\su(1)$}
Let us start the first step of the induction $N-1=1$ which is the case of a single ($-2$)-curve with an $\su(1)$ gauge group.
This is an empty gauge group. We have $G=\SU(2)/\bZ_2$, and we take the coordinate of $\CCt$ as
\beq
w = \frac{1}{2}(z_1^2 +z_1^{-2})
\eeq
where $z_1^2=e^{2\pi \ii  u^1}$ and $u^1=2u_1$. 
Note that $w$ is invariant under the Weyl symmetry $z\mapsto z^{-1}$.
This coordinate $w$ represents the $\CCt=(\bR \times S^1)^2/\bZ_2$ corresponding to $G=\SU(2)/\bZ_2$
as $\bC$ at the level of complex structure of $\CCt$. There is only one massless point which is located at $z_1=z_2=1$ or equivalently $w=1$.

The curve with the $\g_1=\su(1)$ group is always adjacent to a curve having $\g_2=\su(2)$ gauge group \cite{Heckman:2015bfa}, 
so the $\su(1)$ gauge theory must have $\su(2)$ flavor symmetry.
We assume that there are $n_f$ hypermultiplets which are doublet under $\su(2)$, or equivalently $2n_f$ half doublets.
The number $2n_f$ is either $1,2$ or $4$ \cite{Heckman:2015bfa}. The theory at the singular point is actually just 
\Nequals{(2,0)} theory of type $A_1$ with free $2n_f-1$ half doublet hypermultiplets.
The $\su(2)$ flavor symmetry is the same as
the $\SU(2)_L$ of the \Nequals{(2,0)} theory obtained as $\SU(2)_L \times \SU(2)_R \subset \fso(5)_R$ under the decomposition
in terms of \Nequals{(1,0)} supersymmetry.

Now we study the moduli space $\CCt$.
First we consider the region where $z_1 \to 0$ and $w \sim z_1^{-2} \to \infty$ which corresponds to a large tensor multiplet scalar vev.
In 6d we have a term $2 \pi \ii  B_1 \wedge I^1$ with \cite{Ohmori:2014kda} $I ^1= c_2(R) -c_2(F_{\g_2})$, 
where $c_2(F_{\g_2})$ is the second Chern class of the $\g_2=\su(2)$ symmetry.
The effective action in 4d is then
\beq
-\frac{1}{2} \log w (c_2(R) -c_2(F_{\g_2})).
\eeq
Then the monodromy coefficients at $w=\infty$ are $\alpha_{\infty}=-1/2$, $\beta_\infty=0$ and $\gamma_\infty=1/2$.

Next, we consider the monodromy around the point $w=1$ at which new massless degrees of freedom appear.
This is very easy. In fact, $w=1$ is the only singular point in the theory other than $w=\infty$, so the monodromy around $w=1$
is exactly the same as the monodromy around $w=\infty$. We immediately get 
$\alpha_{1}=-1/2$, $\beta_1=0$ and $\gamma_1=1/2$.

Near $w=1$, one can check that the coordinate $w-1$ is identified with the Coulomb operator $\tr \Phi^2$ of the \Nequals4  super Yang-Mills
which is obtained by the $T^2$ compactification of the \Nequals{(2,0)} theory. Thus the $\U(1)_R$ charge is given by $q_1=4$.
Using these facts, we get the difference of the anomaly coefficients between the singular point and a generic point of $\CCt$ as
\beq
\delta n_v=\delta n_h = -q_1 \alpha_1=2, ~~~\delta k = q_1 \gamma_1=2.
\eeq
These reproduce exactly the difference between the anomaly coefficients of $\SU(2)/\bZ_2$ \Nequals4  super Yang-Mills and 
its Coulomb branch $\U(1)$ theory. 
Thus, we have proven the assumption for the induction for $N=2$.

\subsubsection{Recursive steps}
Now let us study the theory \eqref{eq:su1theory} with $G=\SU(N)/\bZ_N$.
We consider a one-dimensional subspace $\CC$ of $\CCt$ on which $z:=z_1=\cdots=z_{N-1}$ and $z_N=z^{-N+1}$.
This branch corresponds to the 6d branch \eqref{eq:1dsub}.
Let us recall the assumption of the induction. The gauge group $\g_{N-1}$ supported on the curve with self-intersection $-\frac{N}{N-1}$
is IR free, and the beta function of this IR free gauge group is $2$.

The coordinate appropriate for $\CC$ is given by 
\beq
w=z^{-N}.
\eeq
This is because there is a residual symmetry $z \to \omega z$ where $\omega^N=1$ due to the division by $\bZ_N$ in $G=\SU(N)/\bZ_N$.
The massless point is located at $w=1$, while there are two regions at infinity, $w \to \infty$ and $w \to 0$.

\paragraph{Monodromy around $w\to\infty$ : }
Let us first consider the region $z \to 0$ or $w \to \infty$. In 6d we have $2\pi \ii  B_N \wedge I^{(N)}$, where 
\beq
I^{(N)}=d'^{N-1} c_2(R)+\frac1{N-1}( N c_2(F_{\g_{N-1}})-(N-1)c_2(F_R)-c_2(F_L)-ic_2(F_{\f_i})).
\eeq
Here $d'^{N-1}$ is determined by \eqref{eq:dprime}, $F_{\g_{N-1}}$ is the gauge field strength of $\g_{N-1}$,
and $F_R$, $F_L$ and $F_{\f_i}$ are the flavor symmetry background fields supported on $X_R$, $X_L$ and $X_i$, respectively.
The coefficients of the gauge and flavor second Chern classes in this equation are determined by the (self-)intersection numbers shown in \eqref{eq:1dsub}.
The 4d effective action is $\log z_{N}^{-1} I^{(N)}$, which is
\beq
\begin{split}
	-\frac{N-1}{N} \log w \biggl( d'^{N-1} c_2(R)+ &\frac{N}{N-1} c_2(F_{\g_{N-1}})\\
	-&c_2(F_R)-\frac{1}{N-1}c_2(F_L) -\frac{i}{N-1}c_2(F_{\f_i}) \biggr).
\end{split}
\eeq
The monodromy coefficients are then 
\begin{equation}
\begin{aligned}
\alpha_{\infty}&=-\frac{N-1}{N} d'^{N-1},&\beta_\infty&=0,&\gamma_{\g_{N-1},\infty}&=-1,\\
\gamma_{R,\infty}&=\frac{N-1}{N} ,&\gamma_{L,\infty}&=\frac{1}{N}, &\gamma_{\f_i,\infty}&=\frac{i}{N}.
\end{aligned}
\end{equation}

\paragraph{Monodromy around $w\to0$ : }
Next let us determine the monodromy at $w=0$. This is in a different Weyl chamber, because this point corresponds to a ``negative volume curve''
in the F-theory realization. By using Weyl transformation, this point is transformed to a positive volume point
where $z_1=z^{-N+1}$ and $z_2=\cdots=z_N=z$ with $z \to \infty$. So we have to do the recursive blow-downs of \eqref{eq:su1theory}
from the right. We get
\beq
\begin{array}{ccccccccc}
 & \su(1)& &  \\
X_L& \frac{N}{N-1}& \vev{\frac{1}{N-1}} & X_R\\
	  &\vev{\frac{N-i}{N-1}}\\
	  &X_i\\
	  &[\f_i]\\
\end{array}\label{eq:1dsub2}
\eeq
The curve of self-intersection $-\frac{N}{N-1}$ now supports the $\su(1)$ gauge group.

Here we have encountered a somewhat confusing situation. 
In 4d, we can smoothly go from the point $w =\infty$ to the point $w=0$.
However, in 6d, the point $w=\infty$ corresponds to \eqref{eq:1dsub} in which the curve of self-intersection $-\frac{N}{N-1}$ supports
the gauge group $\g_{N-1}$,
while the point  $w=0$ corresponds to \eqref{eq:1dsub2} which supports a gauge group $\su(1)$.
How can they be consistent?

The resolution comes from the IR gauge group which appears in the compactification of the 6d theory.
When we blow-down the curves $N-2$ times from the right as in \eqref{eq:1dsub2}, 
we do not have an $\su(1)$ gauge group among the curves which are blown down.
Then, as repeatedly discussed in this paper, we get an IR free gauge group in the compactification of this theory.
On the other hand, when we blow-down the curves $N-2$ times from the left as in \eqref{eq:1dsub}, the $\su(1)$ is blown down and hence
we do not get an IR free gauge group from the compactification of the theory obtained by the blow-downs.
This is because of the assumption of induction.
However, the curve which is not blown down supports a gauge group $\g_{N-1}$ which is now IR free by the assumption of induction.
So the IR free gauge group obtained by the compactification of blow-downs of the rightmost $N-2$ curves must be identified with the
IR free gauge group $\g_{N-1}$ in our assumption of induction. 
This is required by the consistency of our discussion here.

Let us return to the computation of the monodromy at $w =0$. This is obtained from the monodromy at $w=\infty$
by replacing $ c_2(F_{\g_{N-1}}) \to c_2(\su(1))=0$, $R \leftrightarrow L$, $ i \to N-i$ and $d'^{N-1} \to d''^{N-1}$, where $d''^{N-1}$ is obtained by
blow-downs from the right. We get
\begin{equation}
\begin{aligned}
\alpha_{0}&=-\frac{N-1}{N} d''^{N-1},&\beta_0&=0,&\gamma_{\g_{N-1},0}&=0,\\
\gamma_{R,0}&=\frac{1}{N} ,&\gamma_{L,0}&=\frac{N-1}{N} ,&\gamma_{\f_i,0}&=\frac{N-i}{N}.
\end{aligned}
\end{equation}

\paragraph{Monodromy at $w=1$ :}
The monodromy at the massless point $w =1$ is the sum of the monodromy at $w=\infty$ and $w=0$, so we get
\begin{align}
\alpha_{1}&=-\sum_{i=1}^{N-1} d^i,&\beta_1&=0,&\gamma_{\g_{N-1},1}&=-1,
&\gamma_{R,1}&=\gamma_{L,1}=\gamma_{\f_i,1}=1,
\end{align}
where we have used
$
\frac{N-1}{N} (d'^{N-1}+d''^{N-1})=\sum_{i=1}^{N-1} d^i
$ which follows from \eqref{eq:dprime}.

The coordinate $\delta w=w-1$ near $w=1$ is identified as 
\beq
\Phi=\diag(\delta w, \cdots, \delta w, -(N-1)\delta w)
\eeq
where $\Phi$ is the adjoint scalar of $G=\SU(N)/\bZ_N$. We also need the fact that the $\U(1)_R$ is unbroken by the vevs of Higgs branch 
operators by which our theory is higgsed to \Nequals4  super Yang-Mills.
Thus we get the same $\U(1)_R$ charge as that of \Nequals4  super Yang-Mills, and hence the $\U(1)_R$ charge of $\delta w$ is given by $q_1=2$.
Therefore, our final results for the changes of the anomaly coefficients are given by
\beq
\delta n_v=\delta n_h = 2\sum_{i=1}^{N-1} d^i, ~~~\delta k_{\g_{N-1}} = -2,~~~\delta k_R=\delta k_L=\delta k_{\f_i}=2.\label{eq:cchange}
\eeq

\paragraph{Synthesis:}
Now we can prove our claim by induction. The assumption of the induction is that the gauge group $\g_{N-1}$
is IR free with the beta function $2$. At the singular point, we get additional contribution $\delta k_{\g_{N-1}}=-2$
to the beta function. Therefore, the beta function becomes zero and $\U(1)_R$ is free from gauge anomaly.
This implies that the theory is conformal at the singular point.

Next we consider the theory with $N $ replaced by $N+1$. After blowing down $N-2$ curves, we get
\beq
\begin{array}{ccccccccc}
 & & \g_{N-1} & \g_{N} &  \\
X_L& \vev{\frac{1}{N-1}} &\frac{N}{N-1}& 2 & X_R
\end{array}.
\eeq
The gauge group $\g_N$ supported on the curve with self-intersection $-2$ is conformal at this point.
From the point of view of the theory on the curve of self-intersection $-\frac{N}{N-1}$, $\g_N$ is a flavor symmetry.
Now we go to the singular point discussed above. This corresponds in 6d to the situation.
\beq
\begin{array}{ccccccccc}
 & & \g_{N} &  \\
X_L& \vev{\frac{1}{N}} &\frac{N+1}{N}& X_R
\end{array}.
\eeq
Then, the beta function of $\g_N$ is changed by the amount $\delta k_{\g_N} =2$, where we used \eqref{eq:cchange} with $\delta k_R$
replaced by $\delta k_{\g_N} $ in this case. So the gauge group $\g_N$ now has the beta function $2$, completing the induction.
This establishes the fact that a theory containing $\su(1)$ gives an SCFT in 4d.

\subsubsection{Anomaly coefficients}

First, we consider generic points of $\CCt$. Then the 6d theory consists of very Higgsable theories, tensor, hyper, and vector multiplets.
In \cite{Ohmori:2015pua}, the following formulas have been found for very Higgsable theories, tensor, hyper, and vector multiplets. We present the formulas in a slightly simplified but equivalent form than the ones given in \cite{Ohmori:2015pua}.
Let us say that the 6d theory in question has the  anomaly polynomial of the form
\beq
I_\text{matter} \supset  p \frac{c_2(R)p_1(T)}{48}+q \frac{7 p_1(T)^2 - 4 p_2(T)}{5760} + r \frac{p_1(T)^2-4p_2(T)}{192} + s \frac{p_1(T) c_2(F)}{24}.
\eeq
Then, the corresponding 4d anomaly coefficients obtained by the $T^2$ compactification are given as
\beq
n_v=2r-p,~~~n_h-n_v=q,~~~k=2s.\label{eq:4dcentral}
\eeq
This gives the central charges of the theory at the generic point of $\CCt$.
If we let $n^\text{6d}_t$, $h^\text{6d}_h$ and $n^\text{6d}_v$ be the numbers of free tensor, hyper and vector multiplets in 6d, respectively,
$p,q$ and $r$ are given as 
\beq
p=n^\text{6d}_t-n^\text{6d}_v+p_\text{vH}, ~~~q=n^\text{6d}_h-n^\text{6d}_t-n^\text{6d}_v+q_\text{vH},~~~ r=n^\text{6d}_t, 
\eeq
where $p_\text{vH}$ and $q_\text{vH}$ are contributions of very Higgsable matters.

The change of the 4d anomaly coefficient is determined by \eqref{eq:cchange}. In the situation in which $\g_1=\su(1)$,
the left flavor symmetry on $X_L$ is empty, so we only consider the flavor symmetry $\f_i$ and $\g_R$.
After going through the above process $N-1$ times, we get
\beq
\delta n_v=\delta n_h = 2\sum_{i=1}^{N-1} (N-i)d^i, ~~~\delta k_R=2,~~\delta k_{\f_i}=2(N-i).
\label{eq:baseAdelta}
\eeq
Combining with the contribution on the generic point on $\CCt$,
we finally obtain the formulas for the central charges of the theories containing $\su(1)$ to be
\begin{equation}
\begin{aligned}
	 n_h-n_v&=q, &	n_v&=2r-p+2\sum_{i=1}^{N-1}(N-i)d^i,\\
	k_{\f_i}&=2s_{\f_i}+2(N-i),& k_R&=2s_R+2,
\end{aligned}
	\label{eq:baseAformula2}
\end{equation}
which is \eqref{eq:baseAformula} we stated before.
The constants $d^i$ can be determined by the algorithm discussed in detail in \cite{Ohmori:2014kda}.
For example, if the gauge group supported on the $i$-th node is $\su(N_i)$ where $N_1=1$,
then we simply have $d^i=N_i$.

As the simplest example, let us consider the \Nequals{(2,0)} theory of type $A_{N-1}$. In this case, we have $d^i=1$ for all $i$ \cite{Ohmori:2014kda}.
We also have $p=N-1$, $q=0$ and $r=N-1$. Therefore, the above formulas give us $n_v=n_h=N^2-1$, which reproduce the 
anomalies of \Nequals4  super Yang-Mills.
More generally, the theories discussed in Sec.~\ref{sec:withoutIR-A} all fall within the class of theories treated in this subsection, and the resulting 4d SCFTs are class S theories of type $A$ on a torus with a single puncture. We have performed extensive checks that the central charge formulas given above is consistent with what is known about the class S theory of type $A$.

\subsection{Theories without IR free gauge group, type $D$}
Let us now move on to the case with $G=\SO(2N)/\bZ_2$, i.e.~we consider the configurations of the curves of the form  
\begin{equation}
	\begin{array}{ccccccccc}
		[\g_L]& \g_{1} & \cdots &\g_{N-2}&  \g_{N-1} &[\g_R]   \\
		X_L& 2& \cdots &2&2&X_R \\
		   &&&2&  & & \\
		   & &&\g_N\\
		   & &&X_D\\
		   & &&[\g_D]
	\end{array},
	\label{eq:dtype1}
\end{equation}
where $X_D$ is a non-compact flavor curve attaching the $\g_N$ node and supporting flavor $\g_D$.
We denote this theory as $\CT^\text{6d}(\g_1[\g_L],\cdots,\g_{N-2};\g_{N-1}[\g_R],\g_N[\g_D])$.
When all $\g_i$ are $\su(k_i)$ and $\g_{L,R,D}=\su(f_{L,R,D})$ with some $k_i$ and $f_{L,R,D}$,
we further abbreviate this to $\CT^\text{6d}(k_1[f_L],\cdots,k_{N-2};k_{N-1}[f_R],k_N[f_D])$.  

We consider the case where one of $-2$ curves should support an $\su(1)$ or empty gauge group.
In this case the 4d theory  $\CT^\text{4d}(\g_1[\g_L],\cdots,\g_{N-2};\g_{N-1}[\g_R],\g_N[\g_D])$ is superconformal at the most singular point of moduli.  
The possible cases are that one of $\g_1,\g_{N-1}$ or $\g_N$ is $\su(1)$.  
By the symmetry of exchanging the $N-1$-th and $N$-th node, we can assume that $\g_1$ or $\g_{N-1}$ is $\su(1)$.

\paragraph{Coordinate system on $\CCt$ for $G=\SO(2N)/\bZ_2$.}
As for $G=\SU(N)/\bZ_N$, we define the variables $v_i$ ($i=1,\cdots,N$) such that
\begin{equation}
(v_1, v_2,\cdots,v_N)= \sum_i u_i H^i,
\end{equation}
where $H^i=(0,\cdots,0,1,-1,0,\cdots,0)$, $(i=1,\cdots,N-1)$ and $H^N=(0,\cdots 1,1)$ are the standard simple roots of $\SO(2N)$.
We also introduce $\bC^\times \cong \mathbb{CP}^1 \setminus \{0, \infty\}$ variables as $z_i=e^{2\pi \ii  v_i}$.
The Weyl transformations can permute $z_i$ and invert even numbers of $z_i$ at once.
The Weyl chamber in the Cartan torus of $\SO(2N)/\bZ_2$ which corresponds to the positive volume curves in F-theory ${\rm Im}(u^i) \geq 0$ is
\begin{equation}
	|z_1| \leq \cdots \leq |z_N| ,\qquad |z_{N-1}z_N|\leq 1.
	\label{eq:Dchamber}
\end{equation}
The $\bZ_2$ division in $G=\SO(2N)/\bZ_2$ acts as changing the signs of all of the $z_i$ simultaneously as $\{z_i\} \to \{ -z_i\}$.

\paragraph{Shrinking the curves supporting $\g_1, \cdots, \g_{N-1}$.}
From the result of the previous subsection for the $A_{N-1}$ case, we get a 4d SCFT when going to the singular locus of the Coulomb branch where the curves supporting $\g_1,\cdots,\g_{N-1}$ are shrunk while keeping the curve supporting $\g_N$ large.
When $\g_{N-1} = \su(1)$, the anomaly cancellation condition of 6d theory forces $\g_{N-2}=\su(2)$ and $\g_{N}=\su(1)$
except for the case of $N=4$. But in the case that $N=4$ and $\g_{N=4}$ is larger than $\su(1)$, $\g_1$ is always $\su(1)$ and we can simply
exchange the nodes as $\g_1 \leftrightarrow \g_{N=4}$ to get to the situation $\g_N=\su(1)$. Therefore we can assume 
$\g_N=\su(1)$ without loss of generality
if $\g_{N-1}=\su(1)$. Then there is no need to care about the central charge of $\g_N$.
The $\U(1)_R$ suffers from no gauge anomaly when we shrink the curve supporting $\g_N$, and the 4d theory at the most singular point is conformal.

When $\g_1=\su(1)$, the formula \eqref{eq:baseAdelta} tells $\delta k_{\g_N}=4$ by shrinking the $\g_1,\cdots \g_{N-1}$ nodes..
Because $\g_N$ is conformal at the generic point of $\CCt$,
the $\g_N$ gauge group is IR free with $k_{\g_N}=4$ when all the curves but the one supporting $\g_N$ are shrunk.

\paragraph{Shrinking the $\g_N$ node.}
When we shrink the curves supporting $\g_1=\su(1),\g_2,\cdots \g_{N-1}$, we get the configuration
\begin{align}
	\begin{array}{ccccccccc}
		[\g_L]&&\g_N&&[\g_R] \\ 
		X_L&\vev{\frac2N}&\frac4N&\vev{\frac{N-2}N}&X_R\\
		   &&X_D\\
		   &&[\g_D]
	\end{array}.
	\label{eq:Dconfig}
\end{align}
This configuration corresponds to $z_1=z_2=\cdots=z_N=:z$ in $\CCt$.
For the global structure of Lie group $G=\SO(2N)/\bZ_2$, we take $w=z^{-2}$ as a $\bZ_2$ invariant variable.
The singular point is at $w=1$,
and the infinities are the $w=\infty$ and $w=0$.
When $N$ is even, there is a Weyl transformation $z \to z^{-1}$.

The point $w=\infty$ is in the Weyl chamber \eqref{eq:Dchamber} and corresponds to the curve configuration \eqref{eq:Dconfig}.
The Green-Schwartz coupling $2\pi \ii B_N\wedge I^{(N)}$ is
\begin{equation}
	I^{(N)} = d^{(N)}c_2(R)+\frac4N c_2(F_{\g_N})-\frac2N c_2(F_L)-\frac{N-2}N c_2(F_R)-c_2(F_D).
\end{equation}
where $d^{(N)}:=\frac{2}{N}\sum_{i=1}^{N-2}id_i+\frac{N-2}{N}d_{N-1}+d_N$.
Noting that $2 \pi \ii u_N=\frac{1}{2}\tr ( 2 \pi \ii u_i H^i)=\frac N2\log z$, the 4d effective action is
\begin{align}
	-\frac{N}{4}\log w\left(d^{(N)}c_2(R)+\frac{4}{N}c_2(F_{\g_N})-\frac{2}{N}c_2(F_L)
	-\frac{N-2}Nc_2(F_R)-c_2(F_D)\right),
\end{align}
and the monodromy coefficients are
\begin{align}
	\alpha_{\infty} = -\frac N4 d^{(N)},~~\beta_\infty=0,~~\gamma_{g_N,\infty}=-1,~~\gamma_{L,\infty}=\frac12,~~\gamma_{R,\infty}=\frac{N-2}{4},~~\gamma_{D,\infty}=\frac N4.
\end{align}

When $N$ is even, because $w=0$ can be mapped to $w=\infty$ by a Weyl transformation, the monodromy coefficients around $w=0$ and $w=\infty$ are the same. Equivalently, the Weyl invariant coordinate of the  subspace in question is $(w+w^{-1})/2$.
Thus we get
\begin{gather}
	\delta n_v=\delta n_h= 2\sum_{i=1}^{N-2} id_i +(N-2)d_{N-1}+Nd_N,\nonumber
	~~ \delta k_{\g_N}=-4,\\ \delta k_L=2, ~~\delta k_R=N-2,~~\delta k_D=N .
	\label{eq:Ddelnk}
\end{gather}
We used the fact that the $R$ charge of $w-1$  is 2. 

When $N$ is odd, the region $w\sim 0$ is mapped to $z_1=z_2=\cdots=z_{N-1}=z^{-1}$, $z_N=z$ by the Weyl transformation.
This region corresponds to the configuration
\begin{align}
	\begin{array}{ccccccccc}
		[\g_L]&&\g_{N-1}&[\g_R] \\ 
		X_L&\vev{\frac2N}&\frac4N&X_R\\
		&&\vev{\frac{N-2}{N}}\\
		&&X_D\\
		&&[\g_D]
	\end{array}.
	\label{eq:Dconfig2}
\end{align}
Since we can smoothly move from \eqref{eq:Dconfig} to \eqref{eq:Dconfig2},
the two gauge groups $\g_N$ and $\g_{N-1}$ should be the same and they are exchanged in the process.
The monodromy coefficients are
\begin{align}
	\alpha_{0} = -\frac N4 d'^{(N)},~~\beta_0=0,~~\gamma_{g_N,0}=-1,~~\gamma_{L,0}=\frac12,~~\gamma_{R,0}=\frac{N}{4},~~\gamma_{D,0}=\frac {N-2}N.
\end{align}
where $d'^{(N)}=2\sum_{i=1}^{N-2}id_i+(N-2)d_N+Nd_{N-1}$.
Note that the monodromy coefficient $\gamma_{g_N,0}$ is for the gauge group $\g_N$ and not for $\g_{N-1}$ because the $\g_{N-1}$ and
$\g_N$ are exchanged when we move the moduli space. Furthermore, in this case of odd $N$,
$d_N$ is equal to $d_{N-1}$, and the flavor symmetries $\g_R$ and $\g_L$ are empty or $\su(1)$. 
Then, using these facts and also the fact that the R-charge of $w-1$ is $2$, we get essentially the same result as \eqref{eq:Ddelnk}.

Combining, we have checked that $k_{\g_N}=4+\delta k_{\g_N}=0$ after shrinking all the curves and we get a 4d SCFT.

\paragraph{Anomaly coefficients.}
Combining the result \eqref{eq:Ddelnk} and the result for $A_{N-1}$ basis,
we get the following result. In the case of $\g_1=\su(1)$, the anomaly coefficients are given by
\begin{equation}
\begin{aligned}
	n_v&=2r-p+N(2\sum_{i=1}^{N-2}d_i+d_{N-1}+d_{N}),& n_h-n_v&=q,\\
	k_R&=2s_R+N,& k_D&=2s_D+N
\end{aligned}
	\label{eq:baseDformula1}
\end{equation}
In a similar way, one can check that the anomaly coefficients in the case $\g_{N-1}=\su(1)$ is given by
\begin{equation}
\begin{aligned}
	n_v&=2r-p+4\sum_{i=1}^{N-1}id_i,& n_h-n_v=q,\\
	 k_L&=2s_L+4,
\end{aligned}
	\label{eq:baseDformula2}
\end{equation}
where we have used the fact that $d_{N-1}=d_N (=1)$ because $\g_{N-1}=\g_N=\su(1)$, and $[\g_R]$, $[\g_D]$ are empty.

\paragraph{Examples.}
Let us first check the result for the case of \Nequals{(2,0)} theory of $D_N$-type.
We have $d^i=1$ for all $i$ \cite{Ohmori:2014kda}, $p=r=N$ and $q=0$.
Both of the formulas \eqref{eq:baseDformula1} and \eqref{eq:baseDformula2} give the same result, and we get 
$n_h=n_v=2N^2-N=\mathrm{dim} \ \fso(2N)$.

Next let us consider the case when gauge groups are of type $A$ with $\g_i=\su(k_i)$.
As is well known,  the anomaly cancellation condition in this case implies
that $2k_i \geq \sum_j k_j$ where the sum runs through the curves adjacent to the $i$th curve, i.e.\ $\eta^{ij}=1$. 
The possible combinations are then
\begin{itemize}
		\item $\CT^\text{6d}(2[2],2,\cdots2;1,1)$,
		\item $\CT^\text{6d}(1,2,3,\cdots2K-1,2K[1],2K,\cdots2K,2K;K,K)$, $(2K\le N-2)$,
		\item $\CT^\text{6d}(1,2,3,\cdots N-2;(N-1)/2[1],(N-1)/2[1])$ when $N$ is odd,
		\item $\CT^\text{6d}(1,2,3,\cdots N-2;N/2[2],(N-2)/2)$ when $N$ is even.
\end{itemize}

Let us  examine the theory $\CT(1,2,2,\cdots,2;1,1)$. This is the case when $K=1$ in the second line above.
When compactified, the argument of Sec.~\ref{sec:little} tells us that this theory should be equivalent to the minimal $D_N$ conformal matter 
gauged by a $\fso(2N)$ vector multiplet. 
The 6d anomaly polynomial is
$2r-p=4N-9$ and $q=1$ therefore $n_h=4(N^2-N-2)$, $n_v=4N^2-4N-9$. Again, both of \eqref{eq:baseDformula1} and \eqref{eq:baseDformula2}
give the same result because $\g_1=\g_{N-1}=\g_N=\su(1)$.
In 4d, the gauged minimal $D_N$ conformal matter can be realized as the $D_N$ type Class S theory on a torus with one simple puncture.
The anomaly contribution of a $D_N$ type simple-puncture is 
\begin{equation}
	n_h^\text{simple}=4N^2-4N-8,\qquad n_v^\text{simple}=4N^2-4N-9,
\end{equation}
and there is no bulk contributions for the torus. This is a nontrivial check of the argument of Sec.~\ref{sec:little}.
Note that the simple puncture is  the puncture of type $[2N-3,3]$.
A similar computation reveals that 
\begin{itemize}
		\item $\CT^\text{4d}(2[2],2,\cdots2;1,1)$  is $\sS_{D_N}\vev{T^2}\{[N-3,1^3]\}$,
		\item $\CT^\text{4d}(1,2,3,\cdots2K-1,2K[1],2K,\cdots2K,2K;K,K)$ for $2K\le N-2$ is \\ 
		$\sS_{D_N}\vev{T^2}\{[2N-2K-1,2K+1]\}$,
		\item $\CT^\text{4d}(1,2,3,\cdots N-2;(N-1)/2[1],(N-1)/2[1])$ for odd $N$ is 
		$\sS_{D_N}\vev{T^2}\{[N^2]\}$,
		\item $\CT^\text{4d}(1,2,3,\cdots N-2;N/2[2],(N-2)/2)$ for even $N$ is also
		$\sS_{D_N}\vev{T^2}\{[N^2]\}$.
\end{itemize} 

There are more \Nequals{(1,0)} theories that are Higgsable to \Nequals{(2,0)} theories of type $D$
such that some of $\fg_i$ is $\su(1)$.
They arise if  the gauge algebra $\g_{N-2}=\su(2)$ is realized by singular fiber of type $IV$,
or some of $\g_i$ are not $\su$, and can be enumerated using the results of  \cite{Heckman:2015bfa}. 
It would be interesting to identify the corresponding 4d SCFTs. 

\section*{Acknowledgments}
We thank M. Del Zotto for discussions.
KO and HS are partially supported by the Programs for Leading Graduate Schools, MEXT, Japan,
via the Advanced Leading Graduate Course for Photon Science
and via the  Leading Graduate Course for Frontiers of Mathematical Sciences and Physics, respectively. 
KO is also supported by JSPS Research Fellowship for Young Scientists.
YT is  supported in part by JSPS Grant-in-Aid for Scientific Research No. 25870159,
and in part by WPI Initiative, MEXT, Japan at IPMU, the University of Tokyo.
The work of KY is supported in part by DOE Grant No. DE-SC0009988.
 
\newpage
\appendix
 
\section{Field-theoretical inconsistency of some anomaly-free models in 6d}\label{appendix}

In \cite{Heckman:2015bfa} and also previously, it was noticed that the 6d \Nequals{(1,0)} model with the structure $\begin{array}[b]{cc}
\mathfrak{su}(2) & \mathfrak{so}(7)\\
2&2
\end{array}$ with the matter content including $\frac12(\mathbf{2},\mathbf{7})$ 
and
$\begin{array}[b]{cc}
\mathfrak{su}(2) & \mathfrak{so}(8)\\
2&n
\end{array}$ with $n=1,2,3$
are free of both local and global 6d gauge anomalies (and therefore included in the list of \cite{Bhardwaj:2015xxa}), but are not allowed as F-theory models.  In this appendix, we describe that a field-theoretical inconsistency exists in these models, independent of the geometric conditions imposed by F-theory.  The essence is captured by the statement:
\begin{claim}
The 6d model \begin{equation}
\text{a free tensor} + \SU(2) + \text{4 flavors},\label{eq:su2flavor4}
\end{equation} at the superconformal point, only has an $\mathfrak{so}(7)_S$ flavor symmetry under which the original four flavors transform as a spinor representation.
In particular, the $\mathfrak{so}(8)$ symmetry on the generic point on the tensor branch is an accidental enhancement. 
\end{claim}
Given this statement, the two models above are clearly excluded, since $\mathfrak{so}(7)_V$ or $\mathfrak{so}(8)$ under which the original four flavors transform as a vector representation is not a subgroup of $\mathfrak{so}(7)_S$.  Now let us see how this statement can be derived. 

\paragraph{Analysis in 4d:}
In Sec.~\ref{sec:simplestX}, we found the 4d theory obtained by the $T^2$ compactification of this 6d theory at the most singular point is given by \eqref{eq:tikz}.
This is an $\SU(2)_u\times \SU(2)_v$ gauge theory with matters in $\frac12(\mathbf{3},\mathbf{2})$ and $\frac 72(\mathbf{1},\mathbf{2})$. 
Giving the vev to the adjoint scalar of $\SU(2)_u$, we just have $\SU(2)_v$ with four flavors, but as we saw in Sec.~\ref{sec:simplest}, this is the S-dual of the original $\SU(2)$ theory with four flavors that descends from the tensor branch of the 6d theory.
This means that the flavor symmetry $\mathfrak{so}(7)$ which the matter field $\frac 72(\mathbf{1},\mathbf{2})$ carries acts on the original four flavors in the spinor representation~\cite{Seiberg:1994aj}. 

We can also analyze the combined system 
\begin{equation}
\begin{array}[b]{cc}
\mathfrak{su}(2) & \mathfrak{so}(7)\\
2&2
\end{array}
\end{equation} with the matter content \begin{equation}
\frac 12(\mathbf{2},\mathbf{1})+
\frac 12(\mathbf{2},\mathbf{7})+
4(\mathbf{1},\mathbf{8})
\end{equation} directly and see an inconsistency.
We saw in Sec.~\ref{sec:marginal} that the 4d theory on the generic point on $\CCt$ (which is the subset of the Coulomb branch where only the vev for the operators that came from the tensor branch is nonzero) should be a superconformal quiver. 
It is indeed the case, and it was shown in \cite{Chacaltana:2011ze} that it has a description as a class S theory of type $D_4$ on a sphere with the following set of punctures: \begin{equation}
[4^2]_\text{red},\quad
[4^2]_\text{blue},\quad
[5,3],\quad
[5,3],\quad
[3,1^5]
\end{equation} where $[5,3]$ is the simple puncture of the $D_4$ theory.
We notice that there are no three punctures of the same type, and therefore there cannot be the action of the Weyl group of $G=\SU(3)$ that needs to be there, so that the vector multiplets parameterizing  $\CCt$  to enhance to $\SU(3)$. 
In a sense, this is an anomaly of the Weyl group, which should still be present on  $\CCt$. 

\paragraph{Analysis in 5d:}
Using the general analysis in Sec.~\ref{sec:five}, we find that the 5d theory obtained by $S^1$ compactification of the theory \eqref{eq:su2flavor4} is described by 
\begin{itemize}
\item the 5d superconformal theory coming from $\SU(2)$ theory with four flavors
\item where the diagonal $\SU(2)_d$ subgroup of the enhanced $\SU(2)_1\times\SU(2)_2$ flavor symmetry is gauged.
\end{itemize}
As is well known, the 5d $\SU(2)$ theory with four flavors in the strongly-coupled limit has the $E_{4+1}=\SO(10)$ symmetry. The flavor symmetry $\SU(2)_1\times\SU(2)_2$ in question is embedded in $\SO(10)$ in the following manner: \begin{equation}
\SU(2)_1\times\SU(2)_2\times \SO(6)\subset \SO(4)\times \SO(6)\subset \SO(10).
\end{equation} 
Then the $\SU(2)_d \cong \SO(3)_d$ subgroup is embedded in $\SO(10)$ as \begin{equation}
\SO(3)_d \times\SO(7)\subset \SO(10).\label{eq:3+7}
\end{equation}
The enhancement from $\SO(8)$ that acts in the vector representation on the original four flavors to $\SO(10)$ is realized by adding a spinor representation of 
the $\SO(8)$. In \eqref{eq:3+7}, the $\SO(8)$ is broken to $\SO(7)$ in such a way that this spinor is decomposed as $\mathbf{8}=\mathbf{7}+\mathbf{1}$.
By the triality of $\SO(8)$, this means that the $\SO(7)$ flavor symmetry in \eqref{eq:3+7} acts on the original four flavors in the spinor representation.

\paragraph{Analysis in 6d?}  Of course it would be more satisfactory if we can find an inconsistency directly in 6d. We hope to come back to this question in a near future.

\bibliographystyle{ytphys}
%\small\baselineskip=.95\baselineskip
%\let\bbb\bibitem\def\bibitem{\itemsep1pt\bbb}
\bibliography{ref}

\end{document}